\documentclass[12pt]{article}
\usepackage{amsmath}
\usepackage{graphicx,psfrag,epsf,color}
\usepackage{enumerate}
\usepackage{natbib}
\usepackage{float}
\usepackage{multirow}
\usepackage{url} 

\usepackage{todonotes}
\usepackage{soul}

\newcommand{\blind}{1}

\begin{document}
\setstcolor{red}

\def\spacingset#1{\renewcommand{\baselinestretch}%
{#1}\small\normalsize} \spacingset{1}


\if1\blind
{
  \title{\bf Using EEG, SPECT, and Multivariate Resampling Methods to
  Differentiate Between Alzheimer's and other Cognitive Impairments.}
  \author{Arne Bathke\footnote{
    Department of Mathematics, University of Salzburg and
    Department of Statistics, University of Kentucky},
    Sarah Friedrich\footnote{
    Department of Statistics, University of Ulm},
    Frank Konietschke\footnote{
    Department of Statistics, University of Texas at Dallas} , Markus Pauly$^\dag$, \\
    Wolfgang Staffen\footnote{
        Department of Neurology, Christian Doppler Medical Centre and Centre for Cognitive Neuroscience, Paracelsus Medical University Salzburg},
    Nicolas Strobl$^\S$,
    and Yvonne H\"oller$^\S$ \\
     }
  \maketitle
} \fi

\if0\blind
{
  \bigskip
  \bigskip
  \bigskip
  \begin{center}
    {\LARGE\bf Using EEG, SPECT, and Multivariate Resampling Methods to
      Differentiate Between Alzheimer Dementia and other Cognitive Impairments.}
\end{center}
  \medskip
} \fi

\bigskip
\begin{abstract}
The incidence of Alzheimer's disease (AD) and other forms of dementia is increasing in most western countries.
For a precise and early diagnosis, several examination modalities exist, 
among them single-photon emission computed tomography (SPECT) 
and the electroencephalogram (EEG). 
The latter is highly available, free of radiation hazards, and non-invasive. 
Thus, its diagnostic utility regarding different stages of dementia is of great interest in neurological research,
along with the question of whether its utility depends on age or sex of the person being examined.
However, SPECT or EEG measurements are intrinsically multivariate, and 
there has been a shortage of sufficiently general inferential techniques 
for the analysis of multivariate data in factorial designs 
when neither multivariate normality nor equal{ity of} covariance matrices across groups should be assumed. 
We adapt an asymptotic model based (parametric) bootstrap approach to this situation,
demonstrate its ability, and use it for a truly multivariate analysis of the EEG and SPECT measurements,
taking into account demographic factors such as age and sex.
These multivariate results are supplemented by marginal effects bootstrap inference whose theoretical properties 
can be derived analogously to the multivariate methods. 
Both inference approaches can have advantages in particular situations, as illustrated in the data analysis.
\end{abstract}

\noindent%
\vfill

\newpage
\spacingset{1.45} 
\section{Introduction}
\label{sec:intro}
The demographic development in most western countries comes along with a rapidly growing incidence
of dementia \citep{Barnes:2011,Prince:2013}.
Several strategies are being developed to face this challenge, among them early diagnosis, early treatment and,
consequently, prevention of a dementing course \citep{Bateman:2015}.
For an accurate and early diagnosis, a rich variety of examination modalities have been evaluated.
For example, single-photon emission computed tomography (SPECT) 
is a well examined and established tool to differentiate Alzheimer's disease (AD)
from other forms, such as frontotemporal dementia and dementia with Lewy bodies \citep{Yeo:2013}.
While SPECT is considered to be a cheap diagnostic tool, the costs for an electroencephalogram (EEG) are even lower.
The EEG has the additional advantages of being highly available, free of radiation hazards, and non-invasive.
Indeed, the EEG has considerable diagnostic utility in early-onset dementia \citep{Micanovic:2014}.
Especially the extraction of biomarkers from the resting EEG is an easily available, and promising approach \citep{Vecchio:2013}.

\begin{figure}[!h]%
\begin{center}
\includegraphics[width=0.7\textwidth]{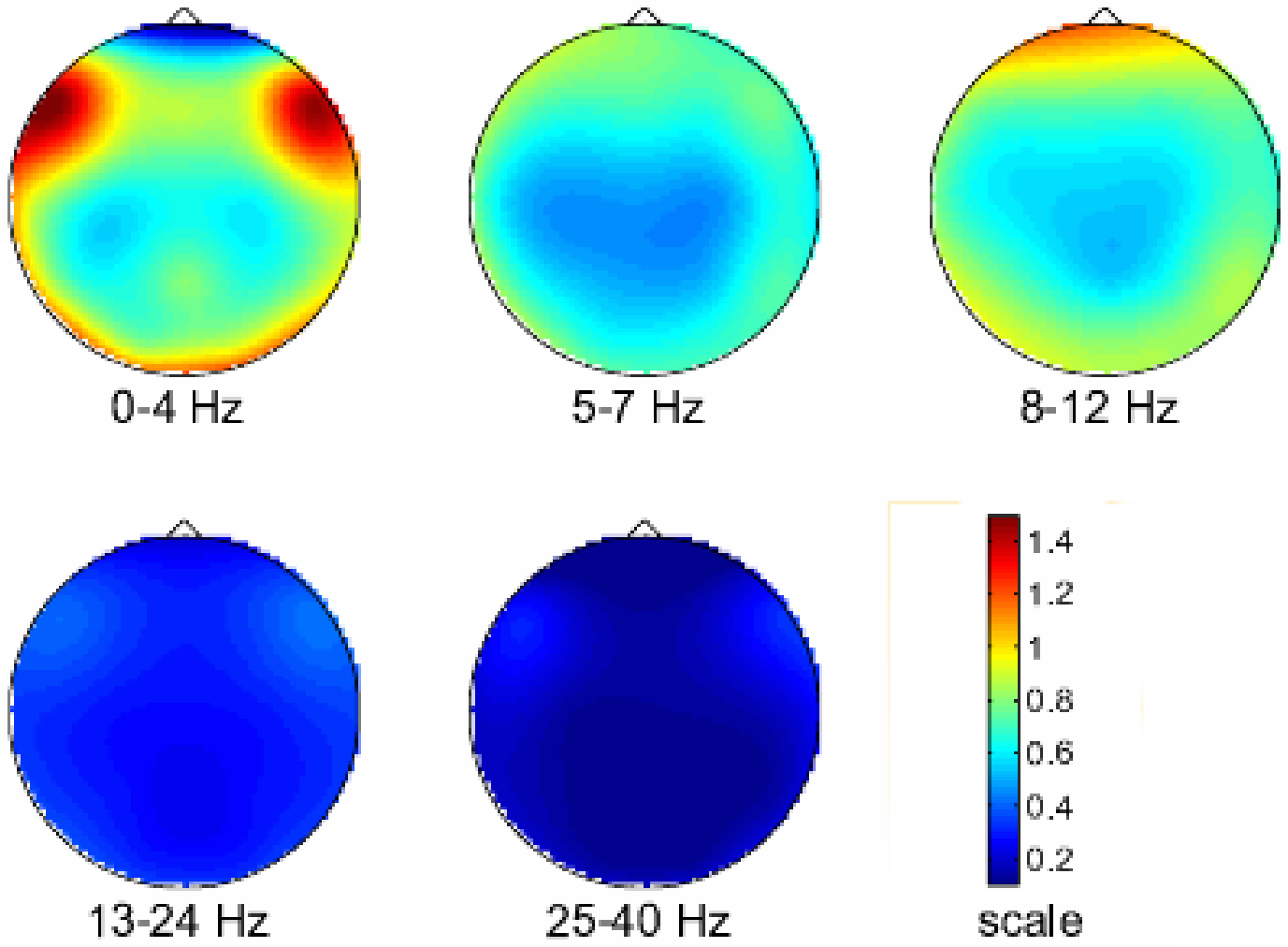}
\end{center}
 \textbf{\refstepcounter{figure}\label{ADFFT} Figure \arabic{figure}.}
 {Topographical maps of EEG activity in $\mu V$ in frequency ranges of interest in a patient sample with AD.}%
\end{figure}

Despite its promise, biomarker research faces very basic problems.
A typical extraction of biomarkers from the EEG results in several markers obtained from different electrode positions
(typically from 21 to 256 channels), possibly being split into different frequency bands (for an example, see Figure \ref{ADFFT}).
Similarly, the quantitative analysis of SPECT requires the evaluation of perfusion values from many possible brain regions of interest.

Considering the above-mentioned comparatively low price and relative ease of EEG recording,
the capacity of EEG is of particular interest in this context.
Could the EEG be applied in order to identify early signs of dementia? 
Can presumably early forms of AD, namely subjective cognitive complaints without clinically significant deficits (SCC) 
and mild cognitive impairment (MCI) be distinguished from AD using EEG alone? Or do we need SPECT to differentiate these different conditions?
Further questions of clinical relevance relate to the effects of demographic factors such as age and sex on EEG and SPECT biomarkers,
and the interactions of age, sex, and diagnosis on the EEG and SPECT features.
In other words, does EEG exhibit a stronger differentiation potential for certain age and sex cohorts,
and which of the cognitive impairment stages can it distinguish then? 
Finally, the structure within the EEG features may exhibit a particular pattern.
There may be differences across the regions, modalities, and types of extracted biomarkers (so-called features), 
or across the spectral distributions,
and{, as stated above,} these within-subjects factors may interact with the between-subjects factors disease status, sex, or age. It is clear that these yield multivariate responses per subject,
obtained in {a factorial design with possible interactions}. 
Traditional analyses of these types of data have often been carried out using essentially univariate techniques
\citep[e.g.,][]{Moretti:2015}.
In such analyses, the multivariate responses were either aggregated into a univariate outcome,
or separate analyses were performed for the different regions, ideally along with some adjustment for multiplicity.
Such a simplifying approach has in large part been driven by the fact 
that appropriate inference {methods} to analyze {broad classes of general}
multivariate data  didn't exist. 
Indeed, the classical multivariate analysis of variance (MANOVA) techniques
(Lawley, 1938; Bartlett, 1939; Wilks, 1946;\nocite{Lawley1938,Bartlett1939,Wilks1946}
Hotelling, 1947; Nanda, 1950; Hotelling, 1951;\nocite{Hotelling1947,Nanda1950,Hotelling1951}
Pillai, 1955; Dempster, 1958, 1960)\nocite{Pillai1955,Dempster1958,Dempster1960}
assume multivariate normal responses with equal covariance matrices across groups. 
However, when covariance matrices do in fact differ and the design is unbalanced (a typical situation in practice) they are known to perform poorly 
(Vallejo and Ato, 2012, Konietschke et
al.,~2015)\nocite{KonietschkeBathkeHarrarPauly2015, VallejoAto2012}. 
For example, in the present paper, we use a SPECT/EEG data set (see Section \ref{sec:verify}), where the observed variance-covariance matrices 
differed greatly between groups (see Table~\ref{CovMatrix} in the Supplementary Material).
Here, the empirical variances for different impairment groups showed up
to almost 50-fold differences (13.84 vs.~0.28 for variable 6 between AD and SCC). 

It is therefore the aim of the present paper to analyze this study with modern resampling methods which do neither assume
multivariate normality nor identical covariance 
matrices across treatment groups. We hereby adopt recently developed procedures that do not suffer from the severe restrictions of classical MANOVA,
while at the same time allowing for a factorial design with basically arbitrary factor structure
(Konietschke et al.,~2015, see also Pauly et al.,~2015).\nocite{PaulyBrunnerKonietschke2015}
The present article is the first attempt to take full advantage of this new methodology,
and to show the strength of modern resampling techniques, combined with the advantages of
a truly multivariate approach to the analysis of data with multiple endpoints.

Additionally, we show how in the special case of repeated measures data 
additional inference can be performed, beyond the questions that are typically addressed using MANOVA methods.
Note that Konietschke et al.~(2015) only considered inference regarding the between-subjects (whole-plot) factors,
their main and simple effects, and possible interactions.
In a typical multivariate data setting, these often constitute the only questions that can 
be feasibly addressed, as multivariate responses may be measured on completely different scales, 
and it may not make sense to make comparisons across different response variables.
If however the variables are commensurate in the sense that such comparisons are meaningful, 
one would be interested in supplementing the inference on between-subjects factors 
by additional inferential results regarding possible within-subjects (sub-plot) factors 
that are structuring the multivariate response vectors,
and by testing for interactions between the two different types of factors (within- and between-subjects).
In the present paper, the results presented by Konietschke et al.~(2015) are generalized to address 
also these additional inferential questions that could arise for commensurate, and structured response vectors.
We have explicitly considered these possibilities in the EEG data analysis,
as they translate to important subject matter questions.

Reviewing the literature on inference methods for multivariate data,
there are very few other approaches which do not assume at least one of either multivariate normality
or covariance matrix equality across groups (or even both).
Among these are the permutation based nonparametric combination methods discussed,
for example, in Pesarin and Salmaso (2010)\nocite{PesarinSalmaso2010} or
Pesarin and Salmaso (2012)\nocite{PesarinSalmaso2012}
(see also Anderson, 2001\nocite{Anderson2001}),
and the fully nonparametric rank-based tests presented in
Bathke and Harrar (2008)\nocite{BathkeHarrar2008},
Bathke et al.~(2008)\nocite{BathkeHarrarMadden2008},
Harrar and Bathke (2008a,b)\nocite{HarrarBathke2008AJMMS,HarrarBathke2008AJMMS},
and Liu et al.~(2011)\nocite{LiuBathkeHarrar2011},
and implemented in the R package {\tt npmv}
(Burchett and Ellis, 2015\nocite{BurchettEllis2015npmv}).
However, these methods are currently limited to the one-way layout,
or to some particular factorial design situations
(Hahn and Salmaso, 2015)\nocite{HahnSalmaso2015}.
Thus, they are not applicable to data from complex factorial designs,
such as those described above.
Also, methodologically, the mentioned articles are not directly comparable to our approach,
as the hypotheses tested are formulated using the distribution functions,
or exchangeability of the observation vectors is postulated.
In contrast, the methods presented in this article test
hypotheses that are formulated using contrasts in terms of mean vectors,
and they do not assume exchangeability.

Other procedures based on testing mean vectors,
but derived under the assumption of multivariate normality,
have been presented for different (one- and two-way) designs,
by Nel and Van der Merwe (1986)\nocite{NelMerwe1986},
Krishnamoorthy and Yu (2004, 2012)\nocite{KrishnamoorthyYu2004,KrishnamoorthyYu2012},
Belloni and Didier (2008)\nocite{BelloniDidier2008},
Giron and Castillo (2010)\nocite{GironCastillo2010},
Krishnamoorthy and Lu (2010)\nocite{KrishnamoorthyLu2010},
Zhang (2011, 2012, 2013)\nocite{Zhang2011two,Zhang2012approximate,Zhang2013tests},
Xu et al.~(2013)\nocite{XuFangAbulaQin2013},
Zhang and Liu (2013)\nocite{ZhangLiu2013},
and Kawasaki and Seo (2015)\nocite{KawasakSeo2015}.

Without the normality assumption, but requiring homogeneous covariance matrices,
Van Aelst and Willems (2011)\nocite{VanAelstWillems2011}
have derived robust one-way MANOVA tests,
which are implemented in the R package {\tt FRB}
(Van Aelst and Willems, 2013\nocite{VanAelstWillems2013}).

Among the several heuristic approaches are a median based MANOVA suggested by
Xu and Cui (2008)\nocite{XuCui2008},
and a multivariate multiple comparison procedure
by Santos and Ferreira (2012)\nocite{SantosFerreira2012}.

Apart from Konietschke et al.~(2015),
the only other mean-based inference method using a multivariate factorial model
without normality or equal covariance matrix assumption is that of
Harrar and Bathke (2012)\nocite{HarrarBathke2012}.
However, due to its design limitations,
it was also not adequate for the analysis of the sample data presented above. 

The paper is organized as follows. In Section~\ref{sec:meth} we state the general factorial models, formulate the hypotheses of interest and explain the corresponding bootstrap test procedure. 
Afterwards Section~\ref{sec:verify} presents the main part of the paper: An extensive statistical analysis of the SPECT/EEG data set within a multivariate 
factorial framework. We then close the paper with a discussion and conclusions. 
We note that simulation results  and additional tables can be found in the supplementary material.

\section{Methods}
\label{sec:meth}
{The data described in Section~\ref{sec:intro} can be described by the multivariate linear model} 
\begin{align} \label{eq:model}
  {\mathbf X}_{ir} &= {\boldsymbol \mu}_i + {\boldsymbol \varepsilon}_{ir},~i=1,\ldots,d,~r=1,\ldots,n_i, \; N=\sum_{i=1}^d n_i,
\end{align}
where the index $i$ represents the treatment group, sample, or, in a factorial design,
the treatment combination,
while $r$ {models} the experimental unit or subject on which $p$-variate observations are being obtained.
In order {to derive asymptotic results} we employ the 
following mild regulatory assumptions:
\begin{itemize} 
\item the error terms ${\boldsymbol \varepsilon}_{i1}, \ldots, {\boldsymbol \varepsilon}_{in_i}$
are independent and identically distributed $p$-dimensional random vectors with
$E({\boldsymbol \varepsilon}_{i1}) = 0$,
$Cov({\boldsymbol \varepsilon}_{i1}) = {\boldsymbol \Sigma}_i > 0$, {and}
$E(\|{\boldsymbol \varepsilon}_{i1}\|^4) <\infty,$ {for}
$i=1,\ldots,d$ {; and}
\item the different sample sizes $n_i$ grow at the same rate, {i.e.}
$n_i/N \to \kappa_i > 0, \ i=1,\ldots,d$ as $N \to \infty$.
\end{itemize}
Note that neither normality of the errors, nor equality of their variance-covariance matrices ${\boldsymbol \Sigma}_i$ is assumed.
The distributions of the error vectors ${\boldsymbol \varepsilon}_{ik}$ may even differ across the groups,
as long as their fourth moments are finite.

The vectors from model (\ref{eq:model}) are aggregated into
${\mathbf X} = ({\mathbf X}_{11}', \ldots, {\mathbf X}_{dn_d}')'$,
${\boldsymbol \mu} = ({\boldsymbol \mu}_1', \ldots, {\boldsymbol \mu}_d')'$, \
${\boldsymbol \mu}_i=(\mu_i^{(1)},\ldots,\mu_i^{(p)})', ~i=1,\ldots,d$, \
and ${\boldsymbol \varepsilon} = ({\boldsymbol \varepsilon}_{11}', \ldots, {\boldsymbol \varepsilon}_{dn_d}')'$, respectively. 
Below we describe the hypotheses of interest and note that all of them can be written as $H_0: {\mathbf T} {\boldsymbol \mu} = {\mathbf 0}$ 
for an adequate (projection) hypothesis matrix ${\mathbf T}$.

\subsection{Multivariate Hypotheses On Between-Subjects Factors and Their Interactions}
\label{sec:multhyp}
By splitting up the index $i$ into different indices, factorial structures are
introduced within the components of ${\boldsymbol \mu}$ or ${\mathbf X}$.
For a complete three-way MANOVA, for example using the three between-subjects factors
age, sex, and diagnosis, 
as suggested by the data set described above,
the index $i$ is split up into three indices
$i=1,\ldots,a$, $j=1,\ldots,b$, and $k=1,\ldots,c$,
each corresponding to one of the factors $A$ (sex), $B$ (diagnosis), and $C$ (age) involved in the study.
Then, for example, ${\boldsymbol \mu} = ({\boldsymbol \mu}_{111}', \ldots, {\boldsymbol \mu}_{abc}')'$,
where the entries ${\boldsymbol \mu}_{ijk}$ are lexicographically ordered, and $d=abc$.

Hypotheses for the different main, simple, and interaction effects
can be formulated using appropriately chosen contrast matrices.
The hypothesis of no main effect $A$ is written as
$H_0(A): \{ \overline{\boldsymbol \mu}_{1\cdot\cdot} = \dots= \overline{\boldsymbol \mu}_{a\cdot\cdot}\}$
= $ \{ ({\mathbf P}_a \otimes \frac{1}{b} {\mathbf J}_b \otimes \frac{1}{c} {\mathbf J}_c
\otimes {\mathbf I}_p)  {\boldsymbol \mu} = {\mathbf 0} \}$,
where $\overline{\boldsymbol \mu}_{i\cdot\cdot}=\frac{1}{bc} \sum_{j=1}^b \sum_{k=1}^c {\boldsymbol \mu}_{ijk}$,
$i=1,\ldots,a$.
Thus, it corresponds to the case
${\mathbf T} = {\mathbf P}_a \otimes \frac{1}{b} {\mathbf J}_b \otimes \frac{1}{c} {\mathbf J}_c
\otimes {\mathbf I}_p$.
Similarly, the hypotheses of no main effect $B$ and $C$ are given by
$H_0(B): \{\overline{\boldsymbol \mu}_{\cdot 1 \cdot} = \dots= \overline{\boldsymbol \mu}_{\cdot b \cdot}\}$ = $  
\{ (\frac{1}{a} {\mathbf J}_a \otimes {\mathbf P}_b \otimes \frac{1}{c} {\mathbf J}_c
\otimes {\mathbf I}_p)  {\boldsymbol \mu} = {\mathbf 0} \}$,
and
$H_0(C): \{ \overline{\boldsymbol \mu}_{\cdot\cdot 1} = \dots= \overline{\boldsymbol \mu}_{\cdot\cdot c}\}$ {=} $ 
\{ (\frac{1}{a} {\mathbf J}_a \otimes \frac{1}{b} {\mathbf J}_b \otimes {\mathbf P}_c
\otimes {\mathbf I}_p)  {\boldsymbol \mu} = {\mathbf 0} \}$, respectively,
with the averages $\overline{\boldsymbol \mu}_{\cdot j \cdot}$ and $\overline{\boldsymbol \mu}_{\cdot\cdot k}$
defined accordingly.
The hypothesis of, for example, no interaction effect between factors $A$ and $B$ can be written as
$H_0(AB): = \{ ({\mathbf P}_a \otimes {\mathbf P}_b \otimes \frac{1}{c} {\mathbf J}_c
\otimes {\mathbf I}_p)  {\boldsymbol \mu} = {\mathbf 0} \}$,
and the other interaction effects are defined analogously.

\subsection{Hypotheses Involving Within-Subjects Factors}
\label{sec:rephyp}
In situations where the response variables are commensurate in the sense
that comparisons between them are meaningful, it is typically of avail to formulate and test hypotheses involving such comparisons.
This can be particularly interesting when the response vector is structured by one or more within-subjects factors.
Splitting up the index $s=1,\ldots,p$ that denotes the response variable facilitates the formulation of
hypotheses involving within-subjects factors.
In the data example, the EEG values on each subject can be considered commensurate, especially after
standardizing each of the response variables.
Also, they are structured by the two factors brain region (we restricted ourselves to temporal, frontal, and central) 
and feature (we considered the features brain rate and complexity).
Denoting region by the index $r=1,\ldots,p_r$ and feature by $s=1,\ldots,p_s$,
each of the six possible combinations is uniquely defined by the index pair $(r,s)$,
suggesting a natural way to split up the index labeling the responses.
This reminds of a repeated measures analysis, 
and indeed the multivariate model (\ref{eq:model}) presented in this paper contains the repeated measures model as a special case.
Note that this is a rather general repeated measures model, without normality assumption, and without the assumption of covariance matrix equality.

For simplicity, assume in the following that, in addition to the two within-subjects factors brain region and feature,
there is only one between-subjects factor present,
whose levels are $i=1,\ldots,a$ (e.g., diagnosis).
The mean vector
${\boldsymbol \mu}_i=(\mu_i^{(1)},\ldots,\mu_i^{(p)})'$
then becomes
${\boldsymbol \mu}_i=(\mu_i^{(11)},\ldots,\mu_i^{(p_r p_s)})', ~i=1,\ldots,a$,
where the entries are again lexicographically ordered.

With these definitions, formulating the corresponding null hypotheses becomes
rather straightforward.
The matrix ${\mathbf I}_p$ used in the previous section simply needs to be replaced
by appropriate choices.
For example, the hypothesis of no main effect of brain region is written as
$H_0(R): \{ ( \frac{1}{a} {\mathbf J}_a \otimes {\mathbf P}_{p_r}
\otimes \frac{1}{p_s} {\mathbf J}_{p_s}) {\boldsymbol \mu} = {\mathbf 0} \}$,
and that of no interaction between diagnosis and brain region by
$H_0(AR): \{ ( {\mathbf P}_a \otimes {\mathbf P}_{p_r}
\otimes \frac{1}{p_s} {\mathbf J}_{p_s}) {\boldsymbol \mu} = {\mathbf 0} \}$.

A major difference between the hypothesis formulation described here and 
the multivariate hypotheses introduced in Section~\ref{sec:multhyp} is the following.
The multivariate equality of two treatments assumes that the treatment means agree in each response,
while the effect considered here only assumes equality of the treatments when averaging across the responses.
Thus, we will refer to the latter as {\em marginal} effect, 
as opposed to the {\em multivariate} effect defined previously.

\subsection{Test Statistics}
As seen in the preceding two subsections, all relevant hypotheses can be written as 
$H_0: {\mathbf T} {\boldsymbol \mu} = {\mathbf 0}$,
with an appropriate choice of the projection hypothesis matrix ${\mathbf T}$.
The corresponding Wald-type test statistic (WTS) is defined as 
\begin{align} \label{wts}
 Q_N({\mathbf T}) =& N \cdot \overline{\mathbf X}_\cdot' 
 {\mathbf T} ( {\mathbf T} \hat{\mathbf V}_N {\mathbf T})^+ {\mathbf T} 
 \overline{\mathbf X}_\cdot~,
\end{align}
where 
$\overline{\mathbf X}_\cdot = (\overline{\mathbf X}_{1\cdot}',\ldots, \overline{\mathbf X}_{d\cdot}')'$,
$\overline{\mathbf X}_{i\cdot} = \frac1{n_i} \sum_{k=1}^{n_i} {\mathbf X}_{ik}$,  and
\begin{align}\label{empcovmat}
 \hat{\mathbf V}_N =& 
 \text{diag}\Big(\frac{N}{n_i} \widehat{\boldsymbol \Sigma}_i: 1\leq i \leq d\Big), \,
 \widehat{\boldsymbol \Sigma}_i = \frac{1}{n_i-1} \sum_{k=1}^{n_i}
 ({\mathbf X}_{ik} - \overline{\mathbf X}_{i\cdot})
 ({\mathbf X}_{ik} - \overline{\mathbf X}_{i\cdot})'.
\end{align}
Here, $(\cdot)^+$ denotes the Moore-Penrose generalized inverse.
Konietschke et al.~(2015) have shown that, under the technical assumptions 
mentioned above, 
$Q_N({\mathbf T})$ has, 
asymptotically, as $N \to \infty$, 
a central $\chi^2$-distribution with degrees of freedom equal to the rank of 
${\mathbf T}$ {under $H_0: {\mathbf T} {\boldsymbol \mu} = {\mathbf 0}$}. 
However, the finite sample performance of the corresponding test is not satisfactory
(see Konietschke et al.,~2015). 
Therefore, different bootstrap methods were proposed by the same authors. 
Out of these, the asymptotic model based bootstrap, 
often referred to as parametric bootstrap, performed the best, 
and it is therefore also considered in the present manuscript.

\subsection{Bootstrap}
\label{sec:boot}

The idea behind the parametric bootstrap approach pursued by Konietschke et al.~(2015) 
originates from an application of the multivariate central limit theorem. 
In particular, we have for any $i=1,\ldots,d$ that 
$\sqrt{n_i} (\overline{\mathbf X}_{i\cdot}-{\boldsymbol \mu}_i)$ is asymptotically 
normal with mean zero and covariance matrix ${\boldsymbol \Sigma}_i$. 
Thus, for approximation purposes, it seems reasonable to replace 
the original i.i.d.~observation vectors, 
 ${\mathbf X}_{i1}, \dots, {\mathbf X}_{in_i}$, 
 in the resample by i.i.d.~parametric bootstrap vectors generated from the estimated limit distribution, 
that is, by 
$$
{\mathbf X}_{i1}^*, \dots, {\mathbf X}_{in_i}^* \stackrel{i.i.d.}{\sim} N({\boldsymbol 0}, \widehat{{\boldsymbol\Sigma}}_i)
$$
for each $i=1,\dots, d$. Recalculating the Wald-type test statistic in \eqref{wts} with 
the variables ${\mathbf X}_{i1}^*, \dots, {\mathbf X}_{in_i}^*$ yields 
$Q_N^*(\mathbf{T})$, the parametric bootstrap version of the WTS. 
The conditional $(1-\alpha)$-quantiles from its distribution, say $c^*(\alpha)$, 
are then used as critical values, 
resulting in the bootstrap test $\varphi_N^* = \mathbf{I}\{Q_N({\mathbf T}) > c^*(\alpha)\}$ for the null hypothesis  $H_0: {\mathbf T} {\boldsymbol \mu} = {\mathbf 0}$.

In their article, Konietschke et al.~(2015) provided simulation results
for different multivariate one- and two-factorial designs.
Our setting differs somewhat due to the more complex structure,
involving within-subjects factors.
In order to investigate whether the present setting allows for use of their method,
we have conducted additional simulations in Section~\ref{sec: simus} of the supplementary material. 

\section{Exemplary Analysis}
\label{sec:verify}
We demonstrate the use of the proposed method by investigating questions 
formulated in a neurological study on cognitive impairments. 
That is, we examine whether EEG- or SPECT-features differentiate SCC, MCI, and AD.

\subsection{Sample}
At the Department of Neurology, University Clinic of Salzburg, 160 patients were diagnosed with either AD, 
MCI, or SCC, 
based on neuropsychological diagnostics for the evaluation of cognitive impairment, 
as well as a thorough neurological examination, excluding any other causes of dementia, such as, for example, vascular or frontotemporal dementia.

\subsection{Examinations and data extraction}
Each of the patients underwent SPECT and EEG. 
From the EEG, brain rate  \citep{Pop-Jordanova:2005} and Hjorth parameters  \citep{Hjorth:1970, Hjorth:1975} 
were calculated from EEG recordings of 21 channels. 
In addition, perfusion values from 46 regions were obtained from the SPECT examinations. 
Our analysis was focused on a number of selected EEG and SPECT variables, 
namely z-scores for brain rate  \citep{Pop-Jordanova:2005} and Hjorth complexity \citep{Hjorth:1970, Hjorth:1975} 
in the EEG (each averaged over frontal, temporal, and central electrode positions across hemispheres in a clinical 10-20 system recording at rest) 
and z-scores of perfusion in the 
medial temporal lobe, lateral temporal lobe, posterior temporal lobe, anterior gyrus cinguli, parietotemporal cortex, and temporal pole 
for SPECT (also for SPECT, averaging was performed over the left and right measurement for each region). 
In addition to standardization, complexity values were multiplied by $-1$ in order to make them more easily comparable to brain rate values:
For brain rate we know that the values decrease with age and pathology, 
while Hjorth complexity values are known to increase with age and pathology.

\subsection{Design configurations}
The three between-subjects factors considered were sex (men vs.~women), 
diagnosis (AD vs.~MCI vs.~SCC), and age ($<70$ vs.~$\geq 70$ years). 
Additionally, we considered the following within-subjects factors structuring the response vector.
For EEG data, we used the selected three brain regions, as well as feature (brain rate or complexity).
For SPECT data, we only used the former (i.e., perfusion values of 6 regions).

We did not consider modality -- that is, EEG vs.~SPECT values -- as another within-subjects factor 
because these variables are not commensurate,
and despite standardization, the method of data acquisition is very different, 
so that the assessed regions can not be matched.

Due to the extremely small number of patients in some groups 
(e.g., only two male patients aged under 70 were diagnosed with AD, see Table~\ref{Freq}), 
we did not consider a layout including all three between-subjects factors but instead restricted our analyses to layouts
with one or two between-subjects factors, as well as one (SPECT) or two (EEG) within-subjects factors.

\begin{table}[ht]
	\centering
	\caption{\it Number of observations for the different factor level combinations.}
	\label{Freq}
	\begin{tabular}{r|r|rrr||r}
		\hline
		sex 	& Age &  AD & MCI & SCC & $\Sigma$ \\ 
		\hline
		M	&	$< 70$ &	2 & 15  &  14 & 31 \\ 
		M	&	$ \geq 70$ & 10    &  12  & 6 & 28\\ \hline
		F	&	$<70$ &  9 &    13 &   29 & 51 \\
		F	&	$\geq 70$ &  15 &   17 & 18  & 50 \\ \hline \hline
		$\Sigma$	&	 & 36   & 57    &  67  & 160\\
		\hline
	\end{tabular}
\end{table}

When using any two of the between-subjects factors, the minimal sample size per factor level combination is 
28 (age and sex), 11 (age and diagnosis), and 12 (diagnosis and sex), see Table~\ref{Freq}. 
Our simulation studies given in Section~\ref{sec: simus} of the supplement have indicated that these are sufficient to ensure 
reasonable performance of the proposed parametric bootstrap procedure. 

Additionally, one-way layouts have been used as basis 
for post-hoc multiple comparison tests regarding the interesting effects.
Here, the minimum cell sample sizes were 78 (age), 36 (diagnosis), and 59 (sex).


For comparison, we present both the results of the classical analysis using the Wald-type statistic 
with a $\chi^2$-approximation and the parametric bootstrap approach described in Section \ref{sec:boot} 
with 10,000 bootstrap runs. 
Each of the analyses was performed using a completely multivariate approach, 
as proposed by Konietschke et al.~(2015), 
and they were supplemented by a marginal effects analysis in which the different responses were considered commensurate,
which allows for the formulation of within-subjects effects. 
In the EEG case, the six responses were considered sub-structured by feature (two levels) and region (three levels),
whereas in the SPECT case, they were simply considered as six levels of the unstructured factor brain region.
Another difference between multivariate and marginal approach is that 
in the former case, possible effects of the between-subjects factors 
are considered in each response variable individually, 
whereas in the latter case they are averaged across the response variables (see also Section~\ref{sec:rephyp}).
This may lead to different results, as demonstrated below.

All the p-values provided in the tables are without correction for multiple testing, 
unless noted specifically.


\subsection{EEG-Results}

The results of the three different multivariate two-way analyses are shown in Table~\ref{twoway_multivariate_EEG}. 
There were clear multivariate effects of diagnosis, while multivariate age and sex effects were only significant
in a design involving exactly these two factors. 
None of the between-subjects factors showed significant interactions. 
Note, that there and throughout the section we will regard results as significant if the p-value is smaller than 5\%.

\begin{table}[H]
	\centering
		\caption{\em Multivariate two-way analysis of EEG data. Factors age ($\ge 70$), diagnosis (AD, MCI, SCC), and sex.
		WTS is the Wald-type statistic approximated by a $\chi^2$-distribution, 
		PBS denotes the asymptotic model based ``parametric'' bootstrap.}
		\label{twoway_multivariate_EEG}
	\begin{tabular}{r|rrr||r}
		\hline
		& Test &  & WTS & PBS \\
		& statistic & df & p-value & p-value \\ 
		\hline
		sex & 15.54 & 6 & 0.0164 & 0.0304 \\ 
		age & 18.69 & 6 & 0.0047 & 0.0101 \\ 
		sex*age & 5.52 & 6 & 0.4792 & 0.5073 \\ 
		\hline
		sex & 12.60 & 6 & 0.0498 & 0.1119 \\ 
		diagnosis & 55.16 & 12 & $<$0.0001 & 0.0005 \\ 
		sex*diagnosis & 9.79 & 12 & 0.6344 & 0.7494 \\ 
		\hline
		diagnosis & 49.44 & 12 & $<$0.0001 & 0.0008 \\ 
		age & 4.84 & 6 & 0.5647 & 0.6182 \\ 
		diagnosis*age & 9.18 & 12 & 0.6876 & 0.7720 \\ 
		\hline
	\end{tabular}
\end{table}

In the marginal analyses incorporating between- and within-subjects factors 
(as described in Section~\ref{sec:rephyp}),
the resulting designs are each four-way layouts using within-subjects factors 
brain region (frontal, central, temporal) and feature (brain rate, complexity),
as well as two of the between-subjects factors age, diagnosis, and sex.
Results using the two between-subjects factors diagnosis and sex
are shown in Table~\ref{fourway_SexDiagnosis_LocationType}.
Those for the other two choices of between-subjects factor pairs (diagnosis and age, sex and age)
are given in the Supplementary Material in
Tables~\ref{fourway_AgeDiagnosis_LocationType} and \ref{fourway_SexAge_LocationType}.

\begin{table}[H]
	\caption{\it Marginal effects analysis. Four-way layouts for EEG data. 
	Between-subjects factors sex and diagnosis (AD, MCI, SCC).
	Within-subjects factors brain region (frontal, central, temporal) and feature (brain rate, complexity).
	WTS stands for the classical Wald-type statistic approximated by a $\chi^2$-distribution, 
	whereas PBS denotes the asymptotic model based ``parametric'' bootstrap procedure.}
	\label{fourway_SexDiagnosis_LocationType}
	\centering
\begin{tabular}{r|ccc||c}
	\hline
	& Test &  & WTS & PBS \\
	{\bf{Effect}}& statistic & df & p-value & p-value \\
	\hline
	sex & 9.97 & 1 & 0.0016 & 0.0045 \\ 
	diagnosis & 42.38 & 2 & $<$0.0001 & $<$0.0001 \\ 
	feature & 0.09 & 1 & 0.7687 & 0.7746 \\ 
	region & 0.07 & 2 & 0.9658 & 0.9657 \\ 
	sex*diagnosis & 3.78 & 2 & 0.1513 & 0.1667 \\ 
	sex*feature & 2.17 & 1 & 0.1410 & 0.1519 \\ 
	sex*region & 0.88 & 2 & 0.6454 & 0.6548 \\ 
	diagnosis*feature & 5.32 & 2 & 0.0701 & 0.0874 \\ 
	diagnosis*region & 6.12 & 4 & 0.1903 & 0.2358 \\ 
	feature*region & 0.65 & 2 & 0.7216 & 0.7310 \\ 
	sex*diagnosis*feature & 1.74 & 2 & 0.4199 & 0.4291 \\ 
	sex*diagnosis*region & 1.53 & 4 & 0.8210 & 0.8375 \\ 
	sex*feature*region & 0.42 & 2 & 0.8095 & 0.8216 \\ 
	diagnosis*feature*region & 7.14 & 4 & 0.1286 & 0.1809 \\ 
	sex*diagnosis*feature*region & 2.27 & 4 & 0.6855 & 0.7131 \\ 
	\hline
\end{tabular}
\end{table}	

The main effects of diagnosis $p<0.0001$ and sex ($p=0.0016$ and $p=0.0045$) 
were always highly significant,
whereas age reached only borderline significance ($p=0.0501$) 
in a design involving sex and age, see Table~\ref{fourway_SexAge_LocationType} in the supplement. 
In that design, there was also a significant interaction effect between age and region ($p=0.0173$).
However, this two-way interaction was not significant when age and diagnosis were used as between-subjects factors.
In the latter case, minimum and average cell sizes were smaller, implying that variances and covariances 
had to be estimated from smaller samples. This may result in lower power in the associated inference.
None of the other main or interaction effects were significant.

Comparing the results from Tables~\ref{twoway_multivariate_EEG} (multivariate analysis) 
and	\ref{fourway_SexDiagnosis_LocationType} (marginal analysis),
there was agreement on the significance of diagnosis, 
and the lack of an interaction effect between diagnosis and sex.
However, there was also a striking difference in the form of a significant ($p=0.0045$) marginal effect of sex,
while the multivariate effect of sex was clearly non-significant ($p=0.1119$). 
This can happen when data are analyzed as multivariate $p$-dimensional, 
but the responses are highly correlated, so that a set of $q<p$ variables carries all the relevant information.
In our case, empirical absolute correlations between the EEG variables ranged from 
0.9266 to 0.9991 for men, and were between 0.8894 and 0.9987 for women.
This contradicts the implicit assumption of a multivariate approach 
that each variable contributes useful information, 
reflected by the degrees of freedom, which equal six for the multivariate test, 
while there is only one degree of freedom in the marginal analysis of the effect of sex.
We emphasize this point here as a caveat regarding the use of multivariate inference methods in general.
They may suffer when the ``relevant'' response space has smaller dimension than $p$.
A correlation analysis between the responses is always advisable, 
and an additional marginal analysis may be useful, where appropriate.
On the other hand, a major advantage of the multivariate approach 
will be demonstrated in the SPECT analysis below.

\begin{figure}[!h]%
\begin{center}
\includegraphics[width=\textwidth]{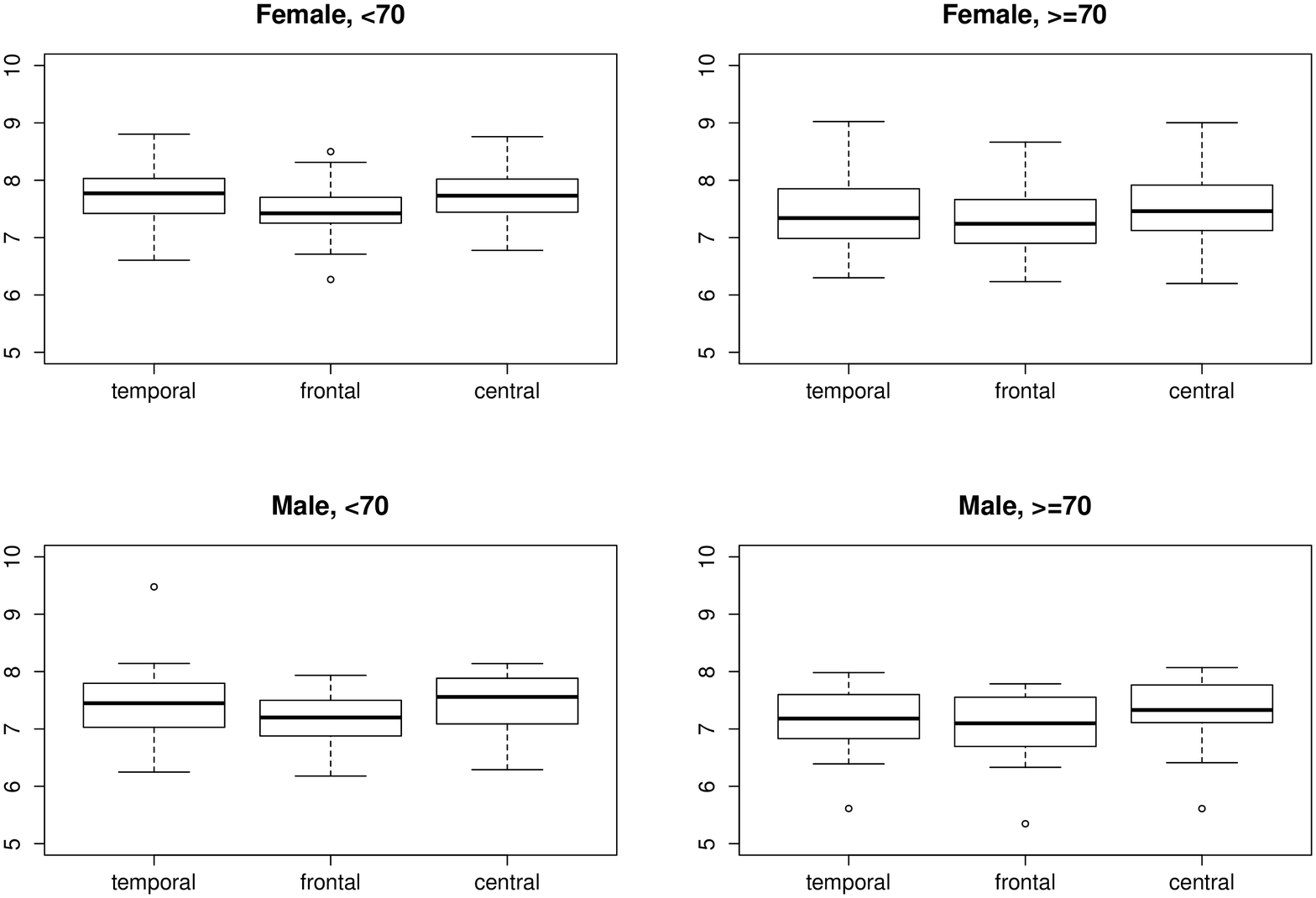}
\end{center}
 \textbf{\refstepcounter{figure}\label{boxplotbrainrate} Figure \arabic{figure}.}
 {Boxplots for EEG {\it brain rate}, separately for sex and age groups.}%
\end{figure}

\begin{figure}[!h]%
\begin{center}
\includegraphics[width=\textwidth]{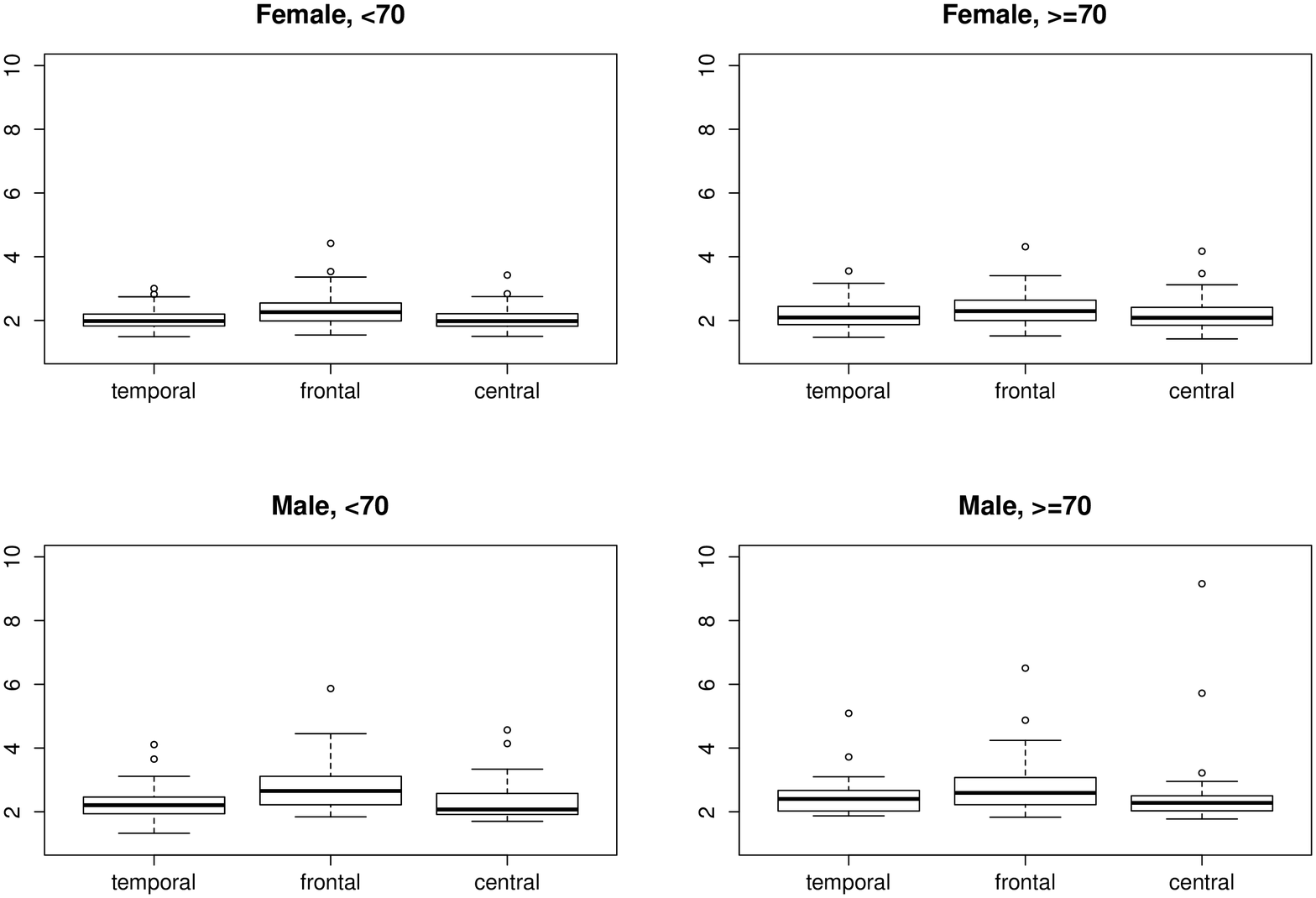}
\end{center}
 \textbf{\refstepcounter{figure}\label{boxplotcomplexity} Figure \arabic{figure}.}
 {Boxplots for EEG {\it complexity}, separately for sex and age groups.}%
\end{figure}

Figures \ref{boxplotbrainrate} and \ref{boxplotcomplexity} 
show box plots of the EEG features brain rate and complexity, respectively.
These aid, for example, in interpreting the possible interaction effects between age and region
that were detected in one of the design configurations. 
A possible explanation is a higher brain rate at temporal than at frontal regions in the younger group 
(especially in female patients), 
while this difference appears to vanish in the older group due to a slowing of the brain rate 
which is most pronounced in the temporal region. 
A similar pattern can be observed for complexity, 
with higher complexity at frontal recording sites in the younger group,
and largely overlapping distributions in the older group,
as complexity increases over temporal and central regions.

Now we assess the effects more closely, starting with the main effect of diagnosis. 
The results for a multivariate pairwise comparison using EEG alone are presented in Table \ref{pwEEG}.
Note that a multiplicity adjustment is not {necessary} for these three pairwise comparisons {by employing the closed testing principle}.
The table shows clearly that EEG could distinguish SCC from AD and MCI, but it could not differentiate between AD and MCI.
\begin{table}[H]
	\caption{\it Pairwise multivariate comparisons of the three different diagnoses using EEG data.
	WTS is the Wald-type statistic using $\chi^2$-distribution quantiles. 
	PBS is the asymptotic model based bootstrap.}
	\label{pwEEG}
	\centering
		\begin{tabular}{c|ccc||c}
			\hline
			& Test &&WTS&PBS\\
		{\bf{ Diagnosis}} 	& statistic & df & p-value & p-value \\
			\hline
			MCI vs.~SCC &	37.17 & 6 &    $<0.0001$   &         $<0.0001$  \\
			AD vs.~SCC &   27.97 & 6 &     0.0001  &          0.0009      \\
			AD vs.~MCI &   5.89 & 6&   0.4352   &          0.5051    \\
			\hline
		\end{tabular}
\end{table}

Next, we investigate which of the variables were driving the detected pairwise distinctions 
(MCI vs.~SCC and AD vs.~SCC).
The results, presented in Tables~\ref{CTP_EEG1} and \ref{CTP_EEG2} in the Supplementary Material, 
demonstrate that all six EEG feature variables were able to significantly distinguish MCI and AD from SCC.
Similar analyses showed that male and female patients were differentiated by the temporal and frontal variables,
while an age effect was only significant in temporal brain rate,
which may have resulted in the interaction effect mentioned above
(Tables~\ref{Sex_EEG} and \ref{Age_EEG} in the Supplementary Material).

Using EEG assessment alone, people with subjective cognitive complaints could be differentiated from patients 
with clinically significant impairments, such as MCI and AD. 
EEG-based differentiation between AD and MCI was not possible.
However, this does not present a challenge in clinical practice 
as this distinction is usually based on neuropsychological tests. 
The differentiation between SCC and MCI could also be done by neuropsychological tests, 
but the results raise a new hope: 
If EEG sensitively differentiates SCC from MCI, it would be worthwhile examining if we can predict 
whether a patient with SCC will convert to MCI after some time. 
Usually, alterations in brain activity occur a long time before behavioral changes may be found. 
Future studies should examine the prognostic value of EEG features. 

Considering the effect of age on the individual EEG variables,
age appeared to impact the brain rate variables, but not complexity
(cf.~Table~\ref{Age_EEG} in the Supplementary Material). 
This is a rather interesting outcome as age is often considered a confounding factor
in EEG based dementia diagnosis \citep{Vecchio:2013}.
Our results seemed to confirm this for brain rate, 
since brain rate decreased both with increasing age and with progress of dementia. 
In contrast however, Hjorth complexity increased with dementia, but was not affected by age. 
This suggests complexity as a promising biomarker for demented patients,
even for age-heterogeneous cohorts.

\subsection{SPECT results}

In the SPECT analysis, we also chose six relevant response variables in order to make 
a fair comparison with the EEG analysis.
Multivariate two-way analyses, analogous to the EEG data analysis from Table~\ref{twoway_multivariate_EEG},
are shown in Table~\ref{twoway_multivariate_SPECT}.
Here, the multivariate effects were significant for each of the between-subjects factors 
diagnosis, age, and sex, and for none of their pairwise interactions.

\begin{table}[H]
	\centering
	\caption{Multivariate two-way analysis of SPECT data. Factors age ($\ge 70$), diagnosis (AD, MCI, SCC), and sex.
			WTS is the Wald-type statistic approximated by a $\chi^2$-distribution, 
			PBS denotes the asymptotic model based ``parametric'' bootstrap.}
    \label{twoway_multivariate_SPECT}
	\begin{tabular}{r|rrr||r}
		\hline
		& Test & & WTS & PBS \\
		& statistic & df & p-value & p-value \\ 
		\hline
		sex & 17.24 & 6 & 0.0084 & 0.0127 \\ 
		age & 21.65 & 6 & 0.0014 & 0.0042 \\ 
		sex*age & 10.81 & 6 & 0.0944 & 0.1176 \\ \hline
		sex & 14.70 & 6 & 0.0227 & 0.0455 \\ 
		diagnosis & 61.55 & 12 & $<$0.0001 & $<$0.0001 \\ 
		sex*diagnosis & 5.73 & 12 & 0.9292 & 0.9517 \\ 
		\hline
		diagnosis & 57.27 & 12 & $<$0.0001 & 0.0001 \\ 
		age & 14.73 & 6 & 0.0225 & 0.0392 \\ 
		diagnosis*age & 11.97 & 12 & 0.4478 & 0.5624 \\ 
		\hline
	\end{tabular}
\end{table}

Contrary to the EEG analysis, the six variables considered here are not structured factorially. 
Instead, in a repeated measures type analysis, we simply regard them as levels
of a within-subjects factor {\em brain region}.
Together with using two of the three between-subjects factors age, sex, and diagnosis 
at a time, we obtain different three-way layouts.
The results for the layout involving sex and diagnosis, as well as brain region,
are shown in Table~\ref{threewaySPECT_genderdiagnosis}.
Those for the other two configurations can be found 
in the Supplementary Material in 
Tables~\ref{threewaySPECT_genderage} and \ref{threewaySPECT_diagnosisage}, respectively.

\begin{table}[ht]
\caption{\it Multivariate three-way analysis of SPECT data. 
	Between-subjects factors sex and diagnosis.
	Within-subjects factor brain region. 
	WTS is the Wald-type statistic approximated by a $\chi^2$-distribution, 
	PBS denotes the asymptotic model based ``parametric'' bootstrap.
	\label{threewaySPECT_genderdiagnosis}}
	\centering
	\begin{tabular}{r|rrr|r}
		\hline
		& Test && WTS & PBS \\
		& statistic & df & p-value & p-value \\ 
		\hline
		sex & 0.01 & 1 & 0.9246 & 0.9231 \\ 
		diagnosis & 51.23 & 2 & $<$0.0001 & $<$0.0001 \\ 
		region & 426.56 & 5 & $<$0.0001 & $<$0.0001 \\ 
		sex*diagnosis & 0.91 & 2 & 0.6333 & 0.6374 \\ 
		sex*region & 14.16 & 5 & 0.0146 & 0.0264 \\ 
		diagnosis*region & 18.31 & 10 & 0.0500 & 0.1119 \\ 
		sex*diagnosis*region & 5.37 & 10 & 0.8651 & 0.8936 \\ 
		\hline
	\end{tabular}
\end{table}

We found significant main effects for the 
between-subjects factor diagnosis and the within-subjects factor region, 
while age was significant only in the layout without diagnosis, 
and perhaps due to the association between these two factors. 
Also, region always interacted significantly with sex, 
while an interaction between age and region was again only significant
in the model not including diagnosis.

The comparison between multivariate and marginal analyses shown in 
Tables~\ref{twoway_multivariate_SPECT} and \ref{threewaySPECT_genderdiagnosis}
reveals an important advantage of the truly multivariate approach.
In the marginal analysis, effects are averaged across the response variables,
whereas the multivariate analysis considers effect contributions of each of the responses individually, 
while taking their correlation into account by construction of the test statistic.
In our case, the effects in the individual SPECT perfusion values did not necessarily point into the same direction.
In fact, the effects of the individual response variables were in part small 
and would not lead to significance using classical variable-wise univariate approaches, 
for example when considering male vs.~female patients
(see Tables~\ref{Sex_SPECT} and \ref{Age_SPECT} in the Supplementary Material).
Furthermore, correlations between the SPECT variables were distributed across much of the $[-1,1]$ interval,
indicating a much lesser degree of multicollinearity between these responses.
That is, each response added information, and only the multivariate analysis took advantage of this information.
It did not make sense to average across the SPECT responses, 
as this led to the masking of some of the available information.

Using a closed testing procedure for pairwise comparisons between the diagnoses,
multivariate inference based on the SPECT values 
detected differences between AD and the other two diagnoses, 
but not between MCI and SCC (see Table~\ref{pwSPECT}).
Differentiation between AD and the other two diagnoses occurred not only using the multivariate tests,
but also individually for each of the six SPECT perfusion values 
(cf.~Table~\ref{CTP_SPECT1} in the Supplementary Material). 

Additional multivariate one-way analyses of the factors sex and age further demonstrated
the advantage of the multivariate approach, as compared to several univariate analyses:
None of the variables individually showed a significant sex effect, 
the multivariate analysis, aggregating information from all variables,
however yielded a clear significance.
Considering several variables in a truly multivariate fashion together 
provides more information than several univariate analyses. 
Note that it is not necessary for the method that the effect directions match for the different variables.
With age, a clear multivariate effect was established, and locally this could be attributed to the
perfusion values from medial and lateral temporal, as well as temporal pole regions.

\begin{table}[H]
	\centering
	\caption{\it Pairwise multivariate comparisons of the three different diagnoses using SPECT perfusion values. WTS is the Wald-type statistic using $\chi^2$-distribution quantiles. 
	PBS is the asymptotic model based bootstrap.}
	\label{pwSPECT}
	\begin{tabular}{r|ccc||c}
		\hline
		& Test && WTS & PBS \\
		{\bf{Diagnosis}} & statistic & df & p-value &  p-value  \\
		\hline
		MCI vs.~SCC   & 5.40 & 6 & 0.4936 & 0.5392 \\ 
		AD vs.~SCC & 58.18 & 6 & $<$0.0001 & $<$0.0001 \\ 
		AD vs.~MCI & 33.74 & 6 & $<$0.0001 & 0.0002 \\ 
		\hline
	\end{tabular}
\end{table}

\subsection{Discussion of the Example}
We found that EEG features differentiated SCC from MCI and SCC from AD. 
There was no interaction effect of the factor diagnosis (AD, MCI, SCC) 
with any of the other between-subjects factors age and sex, 
or with the within-subjects factors feature and region. 
Most important is the absence of a diagnosis $\times$ age interaction.
The assessed EEG features seemed to be robust against normal aging effects. 
We hypothesize therefore that these features could detect symptoms of dementia 
without the confounding effects of age. 
However, we found indications for an interaction effect between age and region.
Specifically, temporal and frontal regions may play a significant role in aging. 
In contrast to EEG, SPECT perfusion values were able to differentiate AD from the other patient groups, 
so that EEG and SPECT perfectly complemented one another. 
The perfusion in most regions was affected by age, which might again reflect normal aging processes. 
Also for the SPECT variables, we did not find a diagnosis $\times$ age interaction.

Finally, we found interesting effects of sex in this clinical sample. 
These appeared more prominent in the EEG than in SPECT measurements. 
Generally, healthy women show a higher amplitude than healthy men in the resting EEG  
\citep{Wada:1994}, 
and higher coherence values, especially for interhemispheric connections in the 
delta, theta, and beta range \citep{Wada:1996}. 
Differences in brain rate and complexity may be based on the same factual differences, 
but altered brain patterns in the demented population should be examined in detail for sex differences. 
We suggest that future studies should directly compare demented populations with healthy participants, 
in order to disentangle disease related processes that may differ between women and men 
from normal sex differences in EEG characteristics, and possibly also in perfusion values.

\section{Conclusion} \label{sec:conc}

We have evaluated the diagnostic utility of EEG and SPECT measurements for the differentiating
diagnosis of different stages of dementia, including AD. Also, it was of interest whether, for example, 
EEG differentiated better between certain diagnosis groups
than between others, and whether the diagnostic utility differed between males and females, 
or between different age groups.
As EEG and SPECT recordings are taken simultaneously at several regions, 
with EEG values being further distinguished by wave spectrum,
the data presented themselves as multivariate, 
in the setting of a factorial design structure with the demographic
factors age and sex, in addition to diagnosis. 

{The need to apply and develop truly multivariate inference methods in order to be able to appropriately
analyze data sets such as the one described has been recognized and articulated already
in some fields of research.
This pertains in particular to medical trials, but applies certainly well beyond the life sciences.
For example, in the context of traumatic brain injuries (TBI),
where the outcome after TBI is {\em per definitionem} multidimensional,
including neuro-physical disabilities and disturbances in mental functioning,
the IMPACT recommendations \citep{Maas2010IMPACT}
``see a need to explore the feasibility of developing a multidimensional approach to outcome assessment and classification''.
\citet{Maas2010} point out that a ``multidimensional approach to outcome assessment is required'' 
while \citet{Bagiella2010} describe the problem that
``no single measure could capture the multidimensional nature of the outcome'',
and \citet{Margulies2009} point out that important deficits could not be identified
using univariate functional assessment scales.
In other contexts, similar arguments have been made \citep{Vester2014}.
For example, \citet{Whitehead2010} state a
``growing interest, especially for trials in stroke, in combining multiple endpoints,''
while \citet{Huang2009} say that
``Parkinson's disease (PD) impairments are multidimensional,
making it difficult to choose a single primary outcome''.}

Until recently, no valid methodology had been developed for the inferential analysis of
multivariate data from factorial designs, unless equal covariance matrices across groups,
or multivariate normality could be assumed. 
For realistic data applications, typically neither of these assumptions is reasonable,
as was also apparent for the EEG and SPECT data considered.
The methodology pursued here is based on an asymptotic model based ``parametric'' bootstrap
approach whose rather general asymptotic validity and good finite sample performance 
have been demonstrated in a recent article (Konietschke et al.,~2015).
We have extended the methodology by enabling inference not only for between-subjects factors,
as is common for multivariate inference, but also for within-subjects factors and the
interactions of all factors involved. 
This corresponds to a repeated measures approach or profile analysis, and, where applicable,
it substantially extends the scope of the possible inferential analysis. 
In the present data, such an extension is sensible due to the commensurate nature of the responses
within the respective groups of EEG and SPECT variables.
The resulting marginal effects analyses are less influenced by multicollinearity of the responses
than a multivariate approach, and thus they provide useful additional information.

Main findings of the data analysis have confirmed the conjectured differentiation ability of EEG 
for early-onset dementia,
and there were no interactions of diagnosis with any of the demographic between-subjects factors
age and sex. The diagnostic utility appears to remain stable across different age and sex cohorts,
although as a limitation to this conclusion, it should be mentioned that 
the study population consisted of mostly elderly people, about half of them 70 years and older.

EEG is a cheap diagnostic and non-invasive tool. 
The study discussed here has demonstrated the utility of EEG for distinguishing subjective cognitive
complaints from more severe forms of dementia. 
At the same time, the analysis has demonstrated the potential of 
novel resampling-based multivariate methods for factorial designs.

\bigskip
\begin{center}
{\large\bf SUPPLEMENTARY MATERIAL}
\end{center}

\begin{description}

\item[Title:] Simulations results and additional information about empirical covariance matrices as well as several results tables from data analyses
using different design configurations. (pdf)

\end{description}

\section*{Acknowledgments}
This work was partly supported by the German Research Foundation project DFG-PA 2409/3-1.

\bibliographystyle{Chicago}

\bibliography{AlzheimerMANOVA}

\begin{thebibliography}{}

\bibitem[\protect\citeauthoryear{Anderson}{Anderson}{2001}]{Anderson2001}
Anderson, M. (2001).
\newblock A new method for non-parametric multivariate analysis of variance.
\newblock {\em Austral ecology\/}~{\em 26\/}(1), 32--46.

\bibitem[\protect\citeauthoryear{Bagiella, Novack, Ansel, Diaz-Arrastia,
  Dikmen, Hart, and Temkin}{Bagiella et~al.}{2010}]{Bagiella2010}
Bagiella, E., T.~A. Novack, B.~Ansel, R.~Diaz-Arrastia, S.~Dikmen, T.~Hart, and
  N.~Temkin (2010).
\newblock Measuring outcome in traumatic brain injury treatment trials:
  {R}ecommendations from the traumatic brain injury clinical trials network.
\newblock {\em The Journal of Head Trauma Rehabilitation\/}~{\em 25\/}(5),
  375--382.

\bibitem[\protect\citeauthoryear{Barnes and Yaffe}{Barnes and
  Yaffe}{2011}]{Barnes:2011}
Barnes, D. and K.~Yaffe (2011).
\newblock The projected effect of risk factor reduction on alzheimer's disease
  prevalence.
\newblock {\em Lancet Neurol\/}~{\em 10}, 819--828.

\bibitem[\protect\citeauthoryear{Bartlett}{Bartlett}{1939}]{Bartlett1939}
Bartlett, M.~S. (1939).
\newblock A note on tests of significance in multivariate analysis.
\newblock In {\em Mathematical Proceedings of the Cambridge Philosophical
  Society}, Volume~35, pp.\  180--185. Cambridge Univ Press.

\bibitem[\protect\citeauthoryear{Bateman}{Bateman}{2015}]{Bateman:2015}
Bateman, R. (2015).
\newblock Alzheimer's disease and other dementias: advances in 2014.
\newblock {\em Lancet Neurol\/}~{\em 14}, 4--6.

\bibitem[\protect\citeauthoryear{Bathke and Harrar}{Bathke and
  Harrar}{2008}]{BathkeHarrar2008}
Bathke, A.~C. and S.~W. Harrar (2008).
\newblock Nonparametric methods in multivariate factorial designs for large
  number of factor levels.
\newblock {\em Journal of Statistical Planning and Inference\/}~{\em 138\/}(3),
  588--610.

\bibitem[\protect\citeauthoryear{Bathke, Harrar, and Madden}{Bathke
  et~al.}{2008}]{BathkeHarrarMadden2008}
Bathke, A.~C., S.~W. Harrar, and L.~V. Madden (2008).
\newblock How to compare small multivariate samples using nonparametric tests.
\newblock {\em Computational Statistics and Data Analysis\/}~{\em 52\/}(11).

\bibitem[\protect\citeauthoryear{Belloni and Didier}{Belloni and
  Didier}{2008}]{BelloniDidier2008}
Belloni, A. and G.~Didier (2008).
\newblock On the {Behrens--Fisher} problem: A globally convergent algorithm and
  a finite-sample study of the wald, lr and lm tests.
\newblock {\em The Annals of Statistics\/}, 2377--2408.

\bibitem[\protect\citeauthoryear{Burchett and Ellis}{Burchett and
  Ellis}{2015}]{BurchettEllis2015npmv}
Burchett, W.~W. and A.~R. Ellis (2015).
\newblock {\em npmv}.
\newblock R package version 2.3.

\bibitem[\protect\citeauthoryear{Dempster}{Dempster}{1958}]{Dempster1958}
Dempster, A.~P. (1958).
\newblock A high dimensional two sample significance test.
\newblock {\em The Annals of Mathematical Statistics\/}, 995--1010.

\bibitem[\protect\citeauthoryear{Dempster}{Dempster}{1960}]{Dempster1960}
Dempster, A.~P. (1960).
\newblock A significance test for the separation of two highly multivariate
  small samples.
\newblock {\em Biometrics\/}~{\em 16\/}(1), 41--50.

\bibitem[\protect\citeauthoryear{Gir{\'o}n and del Castillo}{Gir{\'o}n and del
  Castillo}{2010}]{GironCastillo2010}
Gir{\'o}n, F. and C.~del Castillo (2010).
\newblock The multivariate {Behrens--Fisher} distribution.
\newblock {\em Journal of Multivariate Analysis\/}~{\em 101\/}(9), 2091--2102.

\bibitem[\protect\citeauthoryear{Hahn and Salmaso}{Hahn and
  Salmaso}{2015}]{HahnSalmaso2015}
Hahn, S. and L.~Salmaso (2015).
\newblock A comparison of different synchronized permutation approaches to
  testing effects in two-level two-factor unbalanced anova designs.
\newblock {\em Statistical Papers\/}, doi: 10.1007/s00362--015--0690--2.

\bibitem[\protect\citeauthoryear{Harrar and Bathke}{Harrar and
  Bathke}{2012}]{HarrarBathke2012}
Harrar, S. and A.~Bathke (2012).
\newblock A modified two-factor multivariate analysis of variance: asymptotics
  and small sample approximations (and erratum).
\newblock {\em Annals of the Institute of Statistical Mathematics\/}~{\em
  64\/}(1 \& 5), 135--165 \& 1087.

\bibitem[\protect\citeauthoryear{Harrar and Bathke}{Harrar and
  Bathke}{2008}]{HarrarBathke2008AJMMS}
Harrar, S.~W. and A.~C. Bathke (2008).
\newblock A nonparametric version of the {Bartlett-Nanda-Pillai} multivariate
  test. {A}symptotics, approximations, and applications.
\newblock {\em American Journal of Mathematical and Management Sciences\/}~{\em
  28\/}(3--4).

\bibitem[\protect\citeauthoryear{Hjorth}{Hjorth}{1970}]{Hjorth:1970}
Hjorth, B. (1970).
\newblock {EEG} analysis based on time domain properties.
\newblock {\em Electroencephalogr Clin Neurophysiol\/}~{\em 29}, 306--310.

\bibitem[\protect\citeauthoryear{Hjorth}{Hjorth}{1975}]{Hjorth:1975}
Hjorth, B. (1975).
\newblock Time domain descriptors and their relation to a particular model for
  generation of {EEG} activity.
\newblock In G.~Dolce and H.~Kunkel (Eds.), {\em {CEAN} Computerized {EEG}
  Analysis}, pp.\  3--8. Gustav Fischer.

\bibitem[\protect\citeauthoryear{Hotelling}{Hotelling}{1947}]{Hotelling1947}
Hotelling, H. (1947).
\newblock Multivariate quality control.
\newblock {\em Techniques of statistical analysis\/}.

\bibitem[\protect\citeauthoryear{Hotelling}{Hotelling}{1951}]{Hotelling1951}
Hotelling, H. (1951).
\newblock A generalized t test and measure of multivariate dispersion.

\bibitem[\protect\citeauthoryear{Huang, Goetz, Woolson, Tilley, Kerr, Palesch,
  Elm, Ravina, Bergmann, and Kieburtz}{Huang et~al.}{2009}]{Huang2009}
Huang, P., C.~G. Goetz, R.~F. Woolson, B.~Tilley, D.~Kerr, Y.~Palesch, J.~Elm,
  B.~Ravina, K.~J. Bergmann, and K.~Kieburtz (2009).
\newblock Using global statistical tests in long-term parkinson's disease
  clinical trials.
\newblock {\em Movement Disorders\/}~{\em 24\/}(12), 1732--1739.

\bibitem[\protect\citeauthoryear{Kawasaki and Seo}{Kawasaki and
  Seo}{2015}]{KawasakSeo2015}
Kawasaki, T. and T.~Seo (2015).
\newblock A two sample test for mean vectors with unequal covariance matrices.
\newblock {\em Communications in Statistics-Simulation and Computation\/}~{\em
  44\/}(7), 1850--1866.

\bibitem[\protect\citeauthoryear{Konietschke, Bathke, Harrar, and
  Pauly}{Konietschke et~al.}{2015}]{KonietschkeBathkeHarrarPauly2015}
Konietschke, F., A.~Bathke, S.~Harrar, and M.~Pauly (2015).
\newblock Parametric and nonparametric bootstrap methods for general {MANOVA}.
\newblock {\em Journal of Multivariate Analysis\/}~{\em 140}, 291--301.

\bibitem[\protect\citeauthoryear{Krishnamoorthy and Lu}{Krishnamoorthy and
  Lu}{2010}]{KrishnamoorthyLu2010}
Krishnamoorthy, K. and F.~Lu (2010).
\newblock A parametric bootstrap solution to the {MANOVA} under
  heteroscedasticity.
\newblock {\em Journal of Statistical Computation and Simulation\/}~{\em
  80\/}(8), 873--887.

\bibitem[\protect\citeauthoryear{Krishnamoorthy and Yu}{Krishnamoorthy and
  Yu}{2004}]{KrishnamoorthyYu2004}
Krishnamoorthy, K. and J.~Yu (2004).
\newblock Modified {Nel and Van der Merwe} test for the multivariate
  {Behrens--Fisher} problem.
\newblock {\em Statistics \& probability letters\/}~{\em 66\/}(2), 161--169.

\bibitem[\protect\citeauthoryear{Krishnamoorthy and Yu}{Krishnamoorthy and
  Yu}{2012}]{KrishnamoorthyYu2012}
Krishnamoorthy, K. and J.~Yu (2012).
\newblock Multivariate {Behrens--Fisher} problem with missing data.
\newblock {\em Journal of Multivariate Analysis\/}~{\em 105\/}(1), 141--150.

\bibitem[\protect\citeauthoryear{Lawley}{Lawley}{1938}]{Lawley1938}
Lawley, D. (1938).
\newblock A generalization of {Fisher's} z test.
\newblock {\em Biometrika\/}~{\em 30\/}(1-2), 180--187.

\bibitem[\protect\citeauthoryear{Liu, Bathke, and Harrar}{Liu
  et~al.}{2011}]{LiuBathkeHarrar2011}
Liu, C., A.~Bathke, and S.~Harrar (2011).
\newblock A nonparametric version of {Wilks'} lambda -- asymptotic results and
  small sample approximations.
\newblock {\em Statistics and Probability Letters\/}~{\em 81}, 1502--1506.

\bibitem[\protect\citeauthoryear{Maas, Roozenbeek, and Manley}{Maas
  et~al.}{2010}]{Maas2010}
Maas, A.~I., B.~Roozenbeek, and G.~T. Manley (2010).
\newblock Clinical trials in traumatic brain injury: past experience and
  current developments.
\newblock {\em Neurotherapeutics\/}~{\em 7\/}(1), 115--126.

\bibitem[\protect\citeauthoryear{Maas, Steyerberg, Marmarou, McHugh, Lingsma,
  Butcher, Lu, Weir, Roozenbeek, and Murray}{Maas
  et~al.}{2010}]{Maas2010IMPACT}
Maas, A.~I., E.~W. Steyerberg, A.~Marmarou, G.~S. McHugh, H.~F. Lingsma,
  I.~Butcher, J.~Lu, J.~Weir, B.~Roozenbeek, and G.~D. Murray (2010).
\newblock Impact recommendations for improving the design and analysis of
  clinical trials in moderate to severe traumatic brain injury.
\newblock {\em Neurotherapeutics\/}~{\em 7\/}(1), 127--134.

\bibitem[\protect\citeauthoryear{Margulies and Hicks}{Margulies and
  Hicks}{2009}]{Margulies2009}
Margulies, S. and R.~Hicks (2009).
\newblock Combination therapies for traumatic brain injury: prospective
  considerations.
\newblock {\em Journal of neurotrauma\/}~{\em 26\/}(6), 925--939.

\bibitem[\protect\citeauthoryear{Micanovic and Pal}{Micanovic and
  Pal}{2014}]{Micanovic:2014}
Micanovic, C. and S.~Pal (2014).
\newblock The diagnostic utility of {EEG} in early-onset dementia: a systematic
  review of the literature with narrative analysis.
\newblock {\em J Neural Transm\/}~{\em 121}, 59--69.

\bibitem[\protect\citeauthoryear{Moretti}{Moretti}{2015}]{Moretti:2015}
Moretti, D. (2015).
\newblock Theta and alpha {EEG} frequency interplay in subjects with mild
  cognitive impairment: evidence from {EEG}, {MRI}, and {SPECT} brain
  modifications.
\newblock {\em Frontiers in Aging Neuroscience\/}~{\em 7}.

\bibitem[\protect\citeauthoryear{Nanda}{Nanda}{1950}]{Nanda1950}
Nanda, D. (1950).
\newblock Distribution of the sum of roots of a determinantal equation under a
  certain condition.
\newblock {\em The Annals of Mathematical Statistics\/}, 432--439.

\bibitem[\protect\citeauthoryear{Nel and Van~der Merwe}{Nel and Van~der
  Merwe}{1986}]{NelMerwe1986}
Nel, D. and C.~Van~der Merwe (1986).
\newblock A solution to the multivariate {Behrens-Fisher} problem.
\newblock {\em Communications in Statistics-Theory and Methods\/}~{\em
  15\/}(12), 3719--3735.

\bibitem[\protect\citeauthoryear{Pauly, Brunner, and Konietschke}{Pauly
  et~al.}{2015}]{PaulyBrunnerKonietschke2015}
Pauly, M., E.~Brunner, and F.~Konietschke (2015).
\newblock Asymptotic permutation tests in general factorial designs.
\newblock {\em Journal of the Royal Statistical Society: Series B (Statistical
  Methodology)\/}~{\em 77\/}(2), 461--473.

\bibitem[\protect\citeauthoryear{Pesarin and Salmaso}{Pesarin and
  Salmaso}{2010}]{PesarinSalmaso2010}
Pesarin, F. and L.~Salmaso (2010).
\newblock {\em Permutation tests for complex data: theory, applications and
  software}.
\newblock John Wiley \& Sons.

\bibitem[\protect\citeauthoryear{Pesarin and Salmaso}{Pesarin and
  Salmaso}{2012}]{PesarinSalmaso2012}
Pesarin, F. and L.~Salmaso (2012).
\newblock A review and some new results on permutation testing for multivariate
  problems.
\newblock {\em Statistics and Computing\/}~{\em 22\/}(2), 639--646.

\bibitem[\protect\citeauthoryear{Pillai}{Pillai}{1955}]{Pillai1955}
Pillai, K. (1955).
\newblock Some new test criteria in multivariate analysis.
\newblock {\em The Annals of Mathematical Statistics\/}, 117--121.

\bibitem[\protect\citeauthoryear{{Pop-Jordanova} and
  {Pop-Jordanova}}{{Pop-Jordanova} and
  {Pop-Jordanova}}{2005}]{Pop-Jordanova:2005}
{Pop-Jordanova}, N. and J.~{Pop-Jordanova} (2005).
\newblock Spectrum-weighted {EEG} frequency (''brainrate'') as a quantitative
  indicator of arousal.
\newblock {\em Contributions, Sec Biol Med Sci, MASA, XXVI\/}~{\em 2}, 35--42.

\bibitem[\protect\citeauthoryear{Prince, Bryce, Albanese, Wimo, Ribeiro, and
  Ferri}{Prince et~al.}{2013}]{Prince:2013}
Prince, M., R.~Bryce, E.~Albanese, A.~Wimo, W.~Ribeiro, and C.~Ferri (2013).
\newblock The global prevalence of dementia: a systematic review and
  metaanalysis.
\newblock {\em Alzheimers Dement\/}~{\em 9}, 63--75.

\bibitem[\protect\citeauthoryear{Santos and Ferreira}{Santos and
  Ferreira}{2012}]{SantosFerreira2012}
Santos, E. and D.~Ferreira (2012).
\newblock Multivariate multiple comparisons by bootstrap and permutation tests.
\newblock {\em Rev. Bras. Biom\/}~{\em 30\/}(3), 381--400.

\bibitem[\protect\citeauthoryear{Vallejo and Ato}{Vallejo and
  Ato}{2012}]{VallejoAto2012}
Vallejo, G. and M.~Ato (2012).
\newblock Robust tests for multivariate factorial designs under
  heteroscedasticity.
\newblock {\em Behavior research methods\/}~{\em 44\/}(2), 471--489.

\bibitem[\protect\citeauthoryear{Van~Aelst and Willems}{Van~Aelst and
  Willems}{2011}]{VanAelstWillems2011}
Van~Aelst, S. and G.~Willems (2011).
\newblock Robust and efficient one-way {MANOVA} tests.
\newblock {\em Journal of the American Statistical Association\/}~{\em
  106\/}(494), 706--718.

\bibitem[\protect\citeauthoryear{Van~Aelst and Willems}{Van~Aelst and
  Willems}{2013}]{VanAelstWillems2013}
Van~Aelst, S. and G.~Willems (2013).
\newblock Fast and robust bootstrap for multivariate inference: the {R} package
  {FRB}.
\newblock {\em Journal of Statistical Software\/}~{\em 53\/}(3), 1--32.

\bibitem[\protect\citeauthoryear{Vecchio, Babiloni, Lizio, Fallani~Fde,
  Blinowska, Verrienti, Frisoni, and Rossini}{Vecchio
  et~al.}{2013}]{Vecchio:2013}
Vecchio, F., C.~Babiloni, R.~Lizio, V.~Fallani~Fde, K.~Blinowska, G.~Verrienti,
  G.~Frisoni, and P.~Rossini (2013).
\newblock Resting state cortical {EEG} rhythms in alzheimer's disease: toward
  {EEG} markers for clinical applications: a review.
\newblock {\em Suppl Clin Neurophysiol\/}~{\em 62}, 223--236.

\bibitem[\protect\citeauthoryear{Vester}{Vester}{2014}]{Vester2014}
Vester, J. (2014, October).
\newblock Multivariate inference methods -- a new start in neurosciences
  clinical research.
\newblock Presentation at Workshop Multivariate Inference Methods with
  Applications, International Biometric Society, German Region, D\"usseldorf.

\bibitem[\protect\citeauthoryear{Wada, Nanbu, Kadoshima, Jiang, Koshino, and
  Hashimoto}{Wada et~al.}{1996}]{Wada:1996}
Wada, Y., Y.~Nanbu, R.~Kadoshima, Z.~Jiang, Y.~Koshino, and T.~Hashimoto
  (1996).
\newblock Interhemispheric {EEG} coherence during photic stimulation: sex
  differences in normal young adults.
\newblock {\em Int J Psychophysiol\/}~{\em 22}, 45--51.

\bibitem[\protect\citeauthoryear{Wada, Takizawa, Jiang, and Yamaguchi}{Wada
  et~al.}{1994}]{Wada:1994}
Wada, Y., Y.~Takizawa, Z.~Jiang, and N.~Yamaguchi (1994).
\newblock Gender differences in quantitative {EEG} at rest and during photic
  stimulation in normal young adults.
\newblock {\em Clin Electroencephalogr\/}~{\em 25}, 81--85.

\bibitem[\protect\citeauthoryear{Whitehead, Branson, and Todd}{Whitehead
  et~al.}{2010}]{Whitehead2010}
Whitehead, J., M.~Branson, and S.~Todd (2010).
\newblock A combined score test for binary and ordinal endpoints from clinical
  trials.
\newblock {\em Statistics in Medicine\/}~{\em 29\/}(5), 521--532.

\bibitem[\protect\citeauthoryear{Wilks}{Wilks}{1946}]{Wilks1946}
Wilks, S.~S. (1946).
\newblock Sample criteria for testing equality of means, equality of variances,
  and equality of covariances in a normal multivariate distribution.
\newblock {\em The Annals of Mathematical Statistics\/}, 257--281.

\bibitem[\protect\citeauthoryear{Xu and Cui}{Xu and Cui}{2008}]{XuCui2008}
Xu, J. and X.~Cui (2008).
\newblock Robustified {MANOVA} with applications in detecting differentially
  expressed genes from oligonucleotide arrays.
\newblock {\em Bioinformatics\/}~{\em 24\/}(8), 1056--1062.

\bibitem[\protect\citeauthoryear{Xu, Yang, Abula, and Qin}{Xu
  et~al.}{2013}]{XuFangAbulaQin2013}
Xu, L.-W., F.-Q. Yang, A.~Abula, and S.~Qin (2013).
\newblock A parametric bootstrap approach for two-way {ANOVA} in presence of
  possible interactions with unequal variances.
\newblock {\em Journal of Multivariate Analysis\/}~{\em 115}, 172--180.

\bibitem[\protect\citeauthoryear{Yeo, Lim, Khan, and Pal}{Yeo
  et~al.}{2013}]{Yeo:2013}
Yeo, J., X.~Lim, Z.~Khan, and S.~Pal (2013).
\newblock Systematic review of the diagnostic utility of {SPECT} imaging in
  dementia.
\newblock {\em Eu Arch Psychiatry Clin Neurosci\/}~{\em 263}, 539--552.

\bibitem[\protect\citeauthoryear{Zhang}{Zhang}{2011}]{Zhang2011two}
Zhang, J.-T. (2011).
\newblock Two-way {MANOVA} with unequal cell sizes and unequal cell covariance
  matrices.
\newblock {\em Technometrics\/}~{\em 53\/}(4), 426--439.

\bibitem[\protect\citeauthoryear{Zhang}{Zhang}{2012}]{Zhang2012approximate}
Zhang, J.-T. (2012).
\newblock An approximate {H}otelling {$T^2$}-test for heteroscedastic one-way
  {MANOVA}.
\newblock {\em Open Journal of Statistics\/}~{\em 2}, 1.

\bibitem[\protect\citeauthoryear{Zhang}{Zhang}{2013}]{Zhang2013tests}
Zhang, J.-T. (2013).
\newblock Tests of linear hypotheses in the {ANOVA} under heteroscedasticity.
\newblock {\em International Journal of Advanced Statistics and
  Probability\/}~{\em 1\/}(2), 9--24.

\bibitem[\protect\citeauthoryear{Zhang and Liu}{Zhang and
  Liu}{2013}]{ZhangLiu2013}
Zhang, J.-T. and X.~Liu (2013).
\newblock A modified {B}artlett test for heteroscedastic one-way {MANOVA}.
\newblock {\em Metrika\/}~{\em 76\/}(1), 135--152.

\end{thebibliography}

\newpage
  \section{Supplementary Material for the Article\\
  {\em ``Using EEG, SPECT, and Multivariate Resampling Methods to
  Differentiate Between Alzheimer's and other Cognitive Impairments''}} 

  In this supplementary material we provide simulations results for a design adopted from the discussed data example. 
  Moreover, additional tables with empirical covariance matrices and several results tables from data analyses with 
different design configurations are presented.

\subsection{A Simulation Study}\label{sec: simus}
Type-1 error simulation results for the described Wald-type statistic $Q_N(\mathbf{T})$ as well as the parametric bootstrap tests are given 
in Tables~\ref{Simu2way} and \ref{Simu3way} for different two- and three-way 
factorial designs, respectively. 
In addition we have also implemented the nonparametric bootstrap test described in \citep{KonietschkeBathkeHarrarPauly2015} as another competitor. 
Sample sizes were adopted from the data example (see Table~\ref{Freq}).
In particular, the two-way layout was simulated by neglecting the factor ``age'', 
resulting in the cell sample sizes $\mathbf{n}=(n_{ik}) = (12,27,20,24,30,47)$. 
It can be readily seen that the parametric bootstrap test performed best in controlling 
the nominal level under all considered data distributions. 
The type-1 error simulation results for other sample size configurations yielded similar results and 
were therefore omitted.
	
\begin{table}[H]
	\centering
	\caption{\it Type-1 error simulation results ($\alpha=5\%$) of the Wald-type statistic $Q_N(\mathbf{T})$, 
	nonparametric bootstrap (NPBS), as well as parametric bootstrap tests (PBS) 
	in different two-way factorial designs with sample sizes 
	$\mathbf{n}=(n_{ijk}) = (12,27,20,24,30,47)$ 
	and different underlying data distributions 
	(multivariate normal, double exponential, 
	$\chi^2$-distribution with 20 degrees of freedom,  
	$\chi^2$-distribution with 15 degrees of freedom, 
	$t$-distribution with 7 degrees of freedom).}\label{Simu2way}
	\begin{tabular}{llccc}
Distribution&	Hypothesis	&	WTS	&	NPBS	&	PBS	\\	\hline
Multivariate&	sex	&	0.110	&	0.076	&	0.051	\\
normal      &	diagnosis	&	0.166	&	0.083	&	0.068	\\
            &	sex*diagnosis	&	0.144	&	0.069	&	0.048	\\\hline
Double	    &	sex		&	0.120	&	0.086	&	0.054	\\
exponential &	diagnosis	&	0.173	&	0.080	&	0.058	\\
	        &	sex*diagnosis	&	0.126	&	0.068	&	0.050	\\\hline
$\chi^2(20)$&	sex		&	0.105	&	0.077	&	0.063	\\
            &	diagnosis	&	0.161	&	0.063	&	0.056	\\
	        &	sex*diagnosis&	0.136	&	0.064	&	0.046	\\\hline
$\chi^2(15)$&	sex		&	0.119	&	0.080	&	0.055	\\
            &diagnosis	&	0.137	&	0.057	&	0.047	\\
	        &	sex*diagnosis&	0.155	&	0.088	&	0.059	\\\hline
$t(7)$	    &	sex		&	0.102	&	0.078	&	0.048	\\
	        &	diagnosis&	0.156	&	0.077	&	0.057	\\
	        &	sex*diagnosis&	0.164	&	0.075	&	0.047	\\
\end{tabular}
\end{table}
\newpage
\thispagestyle{empty}
\begingroup
\renewcommand*\baselinestretch{1.1}%
\makeatletter\@currsize\makeatother%

\begin{table}[H]
\centering
\caption{\it Type-1 error simulation results ($\alpha=5\%$) 
of the Wald-type statistic $Q_N(\mathbf{T})$, 
nonparametric bootstrap (NPBS), 
as well as parametric bootstrap tests (PBS) in different three-way factorial 
designs with sample sizes $\mathbf{n}=(n_{ijk}) = (7,15,14,10,12,7,9,13,29,15,17,18)$ 
and different distributions 
(multivariate normal, double exponential, 
$\chi^2$-distribution with 20 degrees of freedom,  
$\chi^2$-distribution with 15 degrees of freedom, 
$t$-distribution with 7 degrees of freedom).}\label{Simu3way}
{\small \begin{tabular}{llccc}\\
Distribution	&	Hypothesis	&	WTS	&	NPBS	&	PBS	\\	\hline
Multivariate    &	sex	&	0.097	&	0.069	&	0.047	\\	
normal	        &	age	&	0.099	&	0.075	&	0.050	\\	
	&	diagnosis	&	0.159	&	0.072	&	0.067	\\	
	&	sex*age	&	0.095	&	0.070	&	0.055	\\	
	&	sex*diagnosis	&	0.114	&	0.061	&	0.044	\\	
	&	age*diagnosis	&	0.151	&	0.074	&	0.055	\\	
	&	sex*age*diagnosis	&	0.119	&	0.071	&	0.051	\\\hline	
Double      	&	sex	&	0.100	&	0.071	&	0.046	\\	
exponential 	&	age	&	0.073	&	0.056	&	0.038	\\	
	&	diagnosis	&	0.151	&	0.070	&	0.053	\\	
	&	sex*age	&	0.091	&	0.066	&	0.045	\\	
	&	sex*diagnosis	&	0.138	&	0.069	&	0.042	\\	
	&	age*diagnosis	&	0.146	&	0.067	&	0.046	\\	
	&	sex*age*diagnosis	&	0.141	&	0.068	&	0.052	\\	\hline	
$\chi^2(20)$   &	sex	&	0.100	&	0.075	&	0.059	\\	
	&	age	&	0.105	&	0.080	&	0.052	\\	
	&	diagnosis	&	0.144	&	0.067	&	0.052	\\	
	&	sex*age	&	0.103	&	0.077	&	0.050	\\	
	&	sex*diagnosis	&	0.138	&	0.063	&	0.046	\\	
	&	age*diagnosis	&	0.134	&	0.057	&	0.043	\\	
	&	sex*age*diagnosis	&	0.147	&	0.067	&	0.047	\\\hline		
$\chi^2(15)$ 	&	sex	&	0.101	&	0.071	&	0.043	\\	
&	age	&	0.094	&	0.066	&	0.057	\\	
	&	diagnosis	&	0.150	&	0.071	&	0.061	\\	
	&	sex*age	&	0.097	&	0.070	&	0.051	\\	
	&	sex*diagnosis	&	0.155	&	0.078	&	0.060	\\	
	&	age*diagnosis	&	0.153	&	0.068	&	0.046	\\	
	&	sex*age*diagnosis	&	0.125	&	0.057	&	0.040	\\\hline		
$t(7)$	&	sex	&	0.095	&	0.073	&	0.051	\\	
	&	age	&	0.095	&	0.063	&	0.048	\\	
	&	diagnosis	&	0.122	&	0.054	&	0.041	\\	
	&	sex*age	&	0.083	&	0.062	&	0.042	\\	
	&	sex*diagnosis	&	0.122	&	0.067	&	0.045	\\	
	&	age*diagnosis	&	0.134	&	0.059	&	0.042	\\	
	&	sex*age*diagnosis	&	0.129	&	0.062	&	0.042	\\	
\end{tabular}}
\end{table}

   \subsection{Additional tables}
   
  \begin{table}[H]
  	\caption{\it Covariance matrices for the three impairment diagnosis groups
  AD, MCI and SCC, calculated for six EEG response variables: 
  three regions and two features of the EEG-signal.
  Variables 1-3 are temporal, frontal, and central values for brain rate, 
  variables 4-6 corresponding values for complexity of the EEG-signal.
  For ease of presentation, the covariance matrices are displayed in tabular form.}
  	\label{CovMatrix}
  	\centering
  	\footnotesize
  	\begin{tabular}{ccccccc}
  		\hline
  		AD	&	1 & 2 & 3 & 4 & 5 & 6 \\
  		\hline
  		&	 5.14 & 5.04 & 4.94& 5.63 & 4.36 & 4.46 \\
  		& 5.04& 6.55& 5.21& 5.74& 5.82&  4.83 \\
  		& 4.94& 5.21& 6.35& 5.39& 4.55 & 6.63 \\
  		& 5.63& 5.74& 5.39& 8.88& 6.92&  6.64 \\
  		& 4.36& 5.82 & 4.55& 6.92& 7.88 & 7.15 \\
  		& 4.46& 4.83& 6.63& 6.64& 7.15& 13.84 \\
  		\hline
  		MCI	&	1 & 2 & 3 & 4 & 5 & 6 \\
  		\hline
  		& 2.10& 1.95& 1.76 & 1.45 & 1.25 & 0.69  \\
  &	 1.95& 2.18 & 1.82 & 1.59 & 1.61 & 0.86  \\
  &	1.76& 1.82& 2.11&  1.41&  1.21 & 1.08  \\
  &	 1.45& 1.59& 1.41 & 2.23&  2.35 & 1.19  \\
  	& 1.25& 1.61& 1.21&  2.35&  2.95 & 1.23  \\
  	& 0.69& 0.86& 1.08 & 1.19 & 1.23 & 1.03 \\
  		
  		\hline
  		SCC	&	1 & 2 & 3 & 4 & 5 & 6  \\
  		\hline
  		& 1.62 & 1.17 & 1.17 & 0.76 & 0.49 & 0.32\\
  		&  1.17 & 1.41 & 1.10 & 0.63 & 0.75 & 0.37\\
  		&  1.17 & 1.10 & 1.26 & 0.54 & 0.39  & 0.41\\
  		&  0.76 & 0.63&  0.54 & 0.64 & 0.53 & 0.30\\
  		&  0.49 & 0.75 & 0.39 & 0.53 & 0.94 & 0.28\\
  		&  0.32&  0.37&  0.41&  0.30&  0.28&  0.28\\
  		\hline
  	\end{tabular}
  \end{table}

  \begin{table}[H]
  	\caption{\it Marginal effects analysis. Four-way layouts for EEG data.
  	Between-subjects factors age ($< 70$ vs.~$\ge 70$ years) and diagnosis (AD, MCI, SCC).
  	Within-subjects factors brain region (frontal, central, temporal) and feature (brain rate, complexity).
  	WTS is the Wald-type statistic approximated by a $\chi^2$-distribution, 
  	PBS denotes the asymptotic model based ``parametric'' bootstrap.}
  	\label{fourway_AgeDiagnosis_LocationType}
  	\centering
  	\begin{tabular}{r|ccc||c}
  		\hline
  		& Test && WTS & PBS \\
  	{\bf{effect}}	& statistic & df & p-value & p-value\\
  		\hline
  		diagnosis & 41.08 & 2 & $<$0.0001 & $<$0.0001 \\ 
  		age & 0.43 & 1 & 0.5096 & 0.5205 \\ 
  		feature & 0.03 & 1 & 0.8585 & 0.8497 \\ 
  		region & 0.39 & 2 & 0.8242 & 0.8258 \\ 
  		diagnosis*age & 1.23 & 2 & 0.5398 & 0.5512 \\ 
  		diagnosis*feature & 4.44 & 2 & 0.1088 & 0.1188 \\ 
  		diagnosis*region & 3.94 & 4 & 0.4142 & 0.4492 \\ 
  		age*feature & 1.98 & 1 & 0.1590 & 0.1628 \\ 
  		age*region & 1.74 & 2 & 0.4192 & 0.4369 \\ 
  		feature*region & 0.11 & 2 & 0.9453 & 0.9492 \\ 
  		diagnosis*age*feature & 0.50 & 2 & 0.7802 & 0.7770 \\ 
  		diagnosis*age*region & 3.98 & 4 & 0.4094 & 0.4354 \\ 
  		diagnosis*feature*region & 7.58 & 4 & 0.1083 & 0.1384 \\ 
  		age*feature*region & 0.66 & 2 & 0.7174 & 0.7179 \\ 
  		diagnosis*age*feature*region & 0.63 & 4 & 0.9602 & 0.9594 \\ 
  		\hline
  	\end{tabular}
  \end{table}	
  
  \begin{table}[H]
  	\caption{\it Marginal effects analysis. Four-way layouts for EEG data. 
  	Between-subjects factors sex and age ($< 70$ vs.~$\ge 70$ years) .
  	Within-subjects factors brain region (frontal, central, temporal) and feature (brain rate, complexity).
  	WTS is the Wald-type statistic approximated by a $\chi^2$-distribution, 
  	PBS denotes the asymptotic model based ``parametric'' bootstrap.}
  	\label{fourway_SexAge_LocationType}
  	\centering
  \begin{tabular}{r|ccc||c}
  	\hline
  	& Test && WTS & PBS \\
  	{\bf{effect}} & statistic & df & p-value & p-value \\
  	\hline
  	sex & 12.16 & 1 & 0.0005 & 0.0008 \\ 
  	age & 3.98 & 1 & 0.0459 & 0.0501 \\ 
  	feature & 0.05 & 1 & 0.8240 & 0.8267 \\ 
  	region & 0.10 & 2 & 0.9508 & 0.9495 \\ 
  	sex*age & 0.02 & 1 & 0.8943 & 0.8951 \\ 
  	sex*feature & 0.90 & 1 & 0.3423 & 0.3493 \\ 
  	sex*region & 1.66 & 2 & 0.4369 & 0.4371 \\ 
  	age*feature & 2.50 & 1 & 0.1138 & 0.1187 \\ 
  	age*region & 8.58 & 2 & 0.0137 & 0.0173 \\ 
  	feature*region & 0.03 & 2 & 0.9852 & 0.9884 \\ 
  	sex*age*feature & 1.02 & 1 & 0.3119 & 0.3178 \\ 
  	sex*age*region & 0.32 & 2 & 0.8504 & 0.8515 \\ 
  	sex*feature*region & 0.37 & 2 & 0.8314 & 0.8339 \\ 
  	age*feature*region & 3.11 & 2 & 0.2107 & 0.2247 \\ 
  	sex*age*feature*region & 0.95& 2 & 0.6220 & 0.6152 \\ 
  	\hline
  \end{tabular}
  \end{table}	
  
  \begin{table}[H]
    \caption{\it  Univariate variable-wise comparison of diagnoses MCI and SCC. 
    EEG values from three regions, two features.
    WTS is the Wald-type statistic approximated by a $\chi^2$-distribution, 
    PBS denotes the asymptotic model based ``parametric'' bootstrap.}
  	\label{CTP_EEG1}
  	\centering
  	\begin{tabular}{l|l|ccc||c}
  		\hline
  	  	&& Test && WTS & PBS \\
  		feature & region & statistic & df & p-value & p-value \\
  		\hline
  		brain rate & temporal &	  26.8455 & 1 &   $<0.0001$          &  $<0.0001$    \\
  		& frontal & 22.77 & 1 &          $<0.0001$         &  $<0.0001$      \\
  		& central & 29.02 & 1 &        $<0.0001$          &  $<0.0001$ \\	
  		\hline
  		complexity& temporal & 21.21 & 1 &       $<0.0001$       &   $<0.0001$       \\
  		& frontal & 8.69 & 1 &  0.0032             &  0.0049 \\
  		& central & 24.3789 & 1 &     $<0.0001$      &    $<0.0001$ \\
  		\hline
  	\end{tabular}
  \end{table}
  
  \begin{table}[H]
  	\centering
  	\caption{\it Univariate variable-wise comparison of diagnoses AD and SCC. 
  	EEG values from three regions, two features.
  	WTS is the Wald-type statistic approximated by a $\chi^2$-distribution, 
  	PBS denotes the asymptotic model based ``parametric'' bootstrap.}
  	\label{CTP_EEG2}
  	\begin{tabular}{l|l|ccc||c}
  		\hline
  		&& Test && WTS & PBS \\
  		feature & region  & statistic & df & p-value & p-value \\
  		\hline
  		brain rate& temporal & 25.27 & 1 &	   $<0.0001$    &      $<0.0001$    \\
  		& frontal & 16.24 & 1 &          0.0001      &  0.0001     \\
  		& central  & 23.01& 1 &      $<0.0001$     &       $<0.0001$ \\
  		\hline
  		complexity & temporal & 17.42 & 1 &     $<0.0001$     &      $<0.0001$    \\
  		& frontal & 7.81 & 1 &      0.0052      &    0.0073   \\
  		& central & 10.33 & 1 &    0.0013       &      0.0021 \\
  		\hline
  	\end{tabular}
  \end{table}
  
  \begin{table}[H]
  	\caption{\it Multivariate vs.~univariate variable-wise analysis for factor sex. 
  	EEG values from three regions, two features.
  	WTS is the Wald-type statistic approximated by a $\chi^2$-distribution, 
  	PBS denotes the asymptotic model based ``parametric'' bootstrap.}
  	\label{Sex_EEG}
  	\centering
  	\begin{tabular}{l|l|ccc||c}
  		\hline
  		&& Test && WTS & PBS \\
  		feature & region  & statistic & df & p-value & p-value \\
  		\hline
  		\multicolumn{2}{c|}{multivariate} &    15.38 & 6 & 0.0175 & 0.0284 \\
  		\hline
  		brain rate & temporal &	        7.98 & 1 & 0.0047 & 0.0064     \\
  		& frontal &         11.02 & 1 & 0.0009 & 0.0010    \\
  		& central &    6.14 & 1 & 0.0132 & 0.0124\\
  		\hline
  		complexity& temporal &        11.91 & 1 & 0.0006 & 0.0006   \\
  		& frontal &       12.55 & 1 & 0.0004 & 0.0008  \\
  		& central  &     6.12 & 1 & 0.0133 & 0.0175  \\
  		\hline
  	\end{tabular}
  \end{table}
  
  \begin{table}[H] 
  	\caption{\it Multivariate vs.~univariate variable-wise analysis for factor age. 
  	 EEG values from three regions, two features.
  	 WTS is the Wald-type statistic approximated by a $\chi^2$-distribution, 
  	 PBS denotes the asymptotic model based ``parametric'' bootstrap.}
  	\label{Age_EEG}
  	\centering
  	\begin{tabular}{l|l|ccc||c}
  		\hline
  	    && Test && WTS & PBS \\
  		feature & region & statistic & df & p-value & p-value\\
  		\hline
  		\multicolumn{2}{c|}{multivariate} &   19.94 & 6 & 0.0028 & 0.0048 \\
  		\hline
  		brain rate& temporal &	 10.82 & 1 & 0.0010 & 0.0008    \\
  		& frontal &     4.46 & 1 & 0.0348 & 0.0347    \\
  		& central &      5.52 & 1 & 0.0188 & 0.0214\\
  		\hline
  		complexity& temporal &   3.52 & 1 & 0.0606 & 0.0623  \\
  		& frontal &    0.06 & 1 & 0.8035 & 0.8045   \\
  		& central  &     2.19 & 1 & 0.1390 & 0.1428 \\
  		\hline
  	\end{tabular}
  \end{table}
  
  \begin{table}[H]
  \caption{\it Three-way layout for SPECT data. 
  	Between-subjects factors sex and age.
  	Within-subjects factor brain region. 
  	WTS is the Wald-type statistic approximated by a $\chi^2$-distribution, 
  	PBS denotes the asymptotic model based ``parametric'' bootstrap.
  	\label{threewaySPECT_genderage}}
  	\centering
  	\begin{tabular}{r|rrr|r}
  		\hline
  		& Test && WTS & PBS \\
  		& statistic & df & p-value & p-value \\ 
  		\hline
  		sex & 0.01 & 1 & 0.9258 & 0.9262 \\ 
  		age & 8.22 & 1 & 0.0042 & 0.0061 \\ 
  		region & 507.27 & 5 & $<$0.0001 & $<$0.0001 \\ 
  		sex*age & 0.03 & 1 & 0.8671 & 0.8625 \\ 
  		sex*region & 16.18 & 5 & 0.0063 & 0.0089 \\ 
  		age*region & 14.46 & 5 & 0.0129 & 0.0177 \\ 
  		sex*age*region & 8.12 & 5 & 0.1497 & 0.1773 \\ 
  		\hline
  	\end{tabular}
  \end{table}
  
  \begin{table}[H]
  \caption{\it Three-way layout for SPECT data. 
  	Between-subjects factors diagnosis and age.
  	Within-subjects factor brain region. 
    WTS is the Wald-type statistic approximated by a $\chi^2$-distribution, 
  	PBS denotes the asymptotic model based ``parametric'' bootstrap.
    \label{threewaySPECT_diagnosisage}}
  	\centering
  	\begin{tabular}{r|rrr|r}
  		\hline
  		& Test && WTS & PBS \\
  		& statistic & df & p-value & p-value \\ 
  		\hline
  		diagnosis & 48.34 & 2 & $<$0.0001 & $<$0.0001 \\ 
  		age & 2.99 & 1 & 0.0835 & 0.0933 \\ 
  		region & 546.81 & 5 & $<$0.0001 & $<$0.0001 \\ 
  		diagnosis*age & 1.01 & 2 & 0.6044 & 0.6125 \\ 
  		diagnosis*region & 20.85 & 10 & 0.0221 & 0.0616 \\ 
  		age*region & 8.59 & 5 & 0.1264 & 0.1585 \\ 
  		diagnosis*age*region & 11.83 & 10 & 0.2966 & 0.4044 \\ 
  		\hline
  	\end{tabular}
  \end{table}
  
  \begin{table}[H]
  	\centering
  	\caption{\it Univariate variable-wise analysis for pairwise comparison between
  	AD and each of the other two diagnoses (MCI, SCC), 
  	using SPECT perfusion values.
    WTS is the Wald-type statistic approximated by a $\chi^2$-distribution, 
  	PBS denotes the asymptotic model based ``parametric'' bootstrap.}
  	\label{CTP_SPECT1}
  	\begin{tabular}{lr|ccc||c}
  		\hline
  		&& Test && WTS & PBS \\
  		&& statistic & df & p-value  & p-value  \\
  		\hline
  		{\bf{AD vs.~SCC}} & medial temporal & 39.05 & 1 &	   $<$0.0001       &  $<$0.0001      \\
  		&lateral temporal & 42.95 & 1 &         $<$0.0001     &     $<$0.0001    \\
  		&posterior temporal & 48.61 & 1 &      $<$0.0001     &     $<$0.0001    \\
  		&anterior gyrus & 18.70 & 1 &        $<$0.0001  &   $<$0.0001  \\
  		&parietotemporal & 34.42 & 1 &    $<$0.0001    &      $<$0.0001 \\
  		&temporal pole    & 29.20 & 1 &     $<$0.0001     &         $<$0.0001 \\
  		\hline
  		{\bf{AD vs.~MCI}} &		medial temporal & 17.07 & 1 &	      $<$0.0001  &            0.0001      \\
  		&lateral temporal & 21.56 & 1 &          $<$0.0001 &          0.0001     \\
  		&posterior temporal & 24.52 & 1 &         $<$0.0001    &           $<$0.0001     \\
  		&anterior gyrus     & 14.46 & 1 &  0.0001  &           0.0002  \\
  		&parietotemporal & 20.09 & 1 &     $<$0.0001       &         0.0001 \\
  		&temporal pole  & 14.70 & 1 &       0.0001  &      0.0002 \\
  		\hline
  	\end{tabular}
  \end{table}

  \begin{table}[H]
  	\centering
  	\caption{\it Multivariate vs.~univariate variable-wise analysis for factor sex, 
  	using SPECT perfusion values.
  	WTS is the Wald-type statistic approximated by a $\chi^2$-distribution, 
  	PBS denotes the asymptotic model based ``parametric'' bootstrap.}
  	\label{Sex_SPECT}
  	\begin{tabular}{r|ccc||c}
  		\hline
  		& Test && WTS & PBS \\
  		{\bf women vs. men} 	& statistic & df & p-value & p-value  \\
  		\hline
  		multivariate           &  17.17 & 6 & 0.0087 & 0.0138 \\ 
  		\hline
  		medial temporal 	   & 0.26 & 1 & 0.6130 & 0.6182 \\ 
  		lateral temporal &        0.04 & 1 & 0.8477 & 0.8488 \\ 
  		posterior temporal &     0.30 & 1 & 0.5812 & 0.5817 \\ 
  		anterior gyrus &     2.32 & 1 & 0.1274 & 0.1286 \\ 
  		parietotemporal &       2.95 & 1 & 0.0858 & 0.0856 \\ 
  		temporal pole & 0.09 & 1 & 0.7629 & 0.7632 \\ 
  		\hline
  	\end{tabular}
  \end{table}
  
  \begin{table}[H]
  	\centering
  	\caption{\it Multivariate vs.~univariate variable-wise analysis for factor age, 
  	using SPECT perfusion values.
    WTS is the Wald-type statistic approximated by a $\chi^2$-distribution, 
  	PBS denotes the asymptotic model based ``parametric'' bootstrap.}
  	\label{Age_SPECT}
  	\begin{tabular}{r|ccc||c}
  		\hline
  		& Test && WTS & PBS \\
  		{\bf $<70$ vs. $\geq 70$} 	& statistic & df & p-value  & p-value  \\
  		\hline
  		multivariate            & 23.30 & 6 & 0.0007 & 0.0020 \\ 
  		\hline
  		medial temporal &	     15.36 & 1 & 0.0001 & 0.0005 \\ 
  		lateral temporal &          10.41 & 1 & 0.0013 & 0.0015 \\ 
  		posterior temporal &     5.89 & 1 & 0.0152 & 0.0154 \\ 
  		anterior gyrus &     5.72 & 1 & 0.0168 & 0.0170 \\ 
  		parietotemporal &     0.78 & 1 & 0.3764 & 0.3746 \\ 
  		temporal pole &  7.56 & 1 & 0.0060 & 0.0060 \\ 
  		\hline
  	\end{tabular}
  \end{table}
  
\endgroup
  
\end{document}